\documentclass[12pt]{article}
\usepackage{hyperref}
\usepackage{mathrsfs}
\usepackage{amsmath}
\usepackage{amsfonts}
\usepackage{mathrsfs}
\usepackage{mathabx}
\usepackage{slashed}
\usepackage{epigraph}
\usepackage{fancyhdr}
\usepackage{amssymb}
\usepackage{graphicx}
\usepackage{longtable}
\usepackage{makecell}
\usepackage{amscd}
\usepackage{tabu}
\usepackage{bm}
\usepackage{color}
\usepackage{cprotect}
\usepackage{array}
\usepackage{verbatim}
\usepackage{cite}
\usepackage{subfig}
\usepackage{setspace}
\usepackage{breakurl}
\usepackage[percent]{overpic}
\usepackage{authblk}
\usepackage{multirow}
\usepackage{xspace}
\usepackage{fullpage}

\def\m{\multirow{2}{*}}

\def\tev{\,{\rm TeV}}
\def\gev{\,{\rm GeV}}
\def\ie{{\it i.e.}}
\def\eg{{\it e.g.}}

\def\to{\rightarrow}

\vspace*{-2cm}
\title{The CP-Violating pMSSM}
\date{}
\author{Joshua Berger$\rm ^{a,b}$}
\author{Matthew W. Cahill-Rowley$\rm ^a$}
\author{Diptimoy Ghosh$\rm ^{c,d}$}
\author{\mbox{JoAnne L. Hewett$\rm ^a$}}
\author{Ahmed Ismail$\rm ^{e,f}$}
\author{Thomas G. Rizzo$\rm ^a$}
\affil{$\rm ^a$SLAC National Accelerator Laboratory,\\ 
2575 Sand Hill Road, Menlo Park, CA 94025, USA
\footnote{jberger, mrowley, hewett, rizzo@slac.stanford.edu}}
\affil{$\rm ^b$Department of Physics,\\ 
University of Wisconsin, Madison, WI  53766, USA}
\affil{$\rm ^c$INFN, Sezione di Roma, \\ 
Piazzale A. Moro 2, I-00185 Roma, Italy}
\affil{$\rm ^d$Department of Particle Physics and Astrophysics,\\
Weizmann Institute of Science, Rehovot 7610001, Israel
\footnote{diptimoy.ghosh@weizmann.ac.il}}
\affil{$\rm ^e$Argonne National Laboratory, \\ 
9700 South Cass Avenue, Argonne, IL 60439, USA
\footnote{aismail@anl.gov}}
\affil{$\rm ^f$University of Illinois at Chicago, \\ 
845 West Taylor Street, Chicago, IL 60607, USA}
\begin{document}
\rightline{\vbox{\halign{&#\hfil\cr&SLAC-PUB-15963\cr}}}
{\let\newpage\relax\maketitle}
\vspace*{-8mm}
\begin{abstract}
We investigate the sensitivity of the next generation of flavor-based
low-energy experiments to probe the supersymmetric parameter space in the context of the phenomenological 
MSSM (pMSSM), and examine the complementarity with direct searches for Supersymmetry at the 13 TeV LHC in a quantitative manner.
To this end, we enlarge the previously studied pMSSM parameter space
to include all physical non-zero CP-violating phases, namely those associated with the gaugino mass parameters, Higgsino mass parameter, 
and the tri-linear couplings of the top quark, bottom quark and tau 
lepton. We find that future electric dipole moment and flavor measurements can have 
a strong impact on the viability of these models even if the sparticle spectrum is out of reach of the 13 TeV
LHC. In particular, the lack of positive signals in future low-energy probes would 
exclude values of the phases between ${\cal O}(10^{-2})$ and ${\cal O}(10^{-1})$.  We also find regions of parameter space where
large phases remain allowed due to cancellations.   Most
interestingly, in some rare processes, such as BR($B_s \to \mu^+ \mu^-$ ), we find that 
contributions arising from CP-violating phases can bring the potentially large SUSY contributions into better 
agreement with experiment and Standard Model predictions.

\end{abstract}
\clearpage
\section{Introduction}
\label{sec:introduction}

The virtual effects of new interactions provide an important window of opportunity to probe the 
presence of physics beyond the Standard Model (SM). In particular, measurements of flavor-changing 
and CP-violating processes yield stringent tests of the SM at the quantum 
loop-level~\cite{Hewett:2012ns,Hewett:2014qja,Antonelli:2009ws,Hewett:1996uc,Charles:2004jd,
Ciuchini:2000de}, 
and are potentially sensitive to new 
physics at scales far beyond the reach attainable at colliders in the foreseeable future.  For 
example, constraints on new contributions to the four-quark operators that mediate neutral meson 
mixing place severe bounds on a new physics scale of up to $10^{3-5}$ TeV for unit coupling strength.  
Alternatively, if new interactions are present at the TeV-scale, these limits constrain the 
effective coupling strength to be below $10^{-11}$ to $10^{-5}$ \cite{Hewett:2012ns}.

The power of low-energy measurements in constraining UV-complete theories is best illustrated by 
the strong limit on sfermion masses in Supersymmetry (SUSY) theories
\cite{Buchmueller:2010ai,Bhattacherjee:2010ju,Ghosh:2012dh,Arbey:2012ax,Dighe:2013wfa}. In particular, the 
Charge-Parity (CP)-conserving and CP-violating observables in meson mixing, together 
with constraints on the electric dipole moment of the neutron and electron, exclude TeV-scale 
SUSY breaking parameters if the CP-violating phases and sfermion mixing angles are of order 
1~\cite{Ellis:1981ts,Chang:1998uc,Martin:1997ns,Isidori:2010kg,Altmannshofer:2013lfa}. For example, 
constraints from neutral kaon mixing require squarks to be 
heavier than $\sim$ 500 TeV if the squark mass matrices have an anarchic flavor structure. Clearly, 
a viable model of TeV-scale Supersymmetry requires a specific flavor structure to accommodate the 
data; such structures are in fact present in many specific models of SUSY 
breaking~\cite{Giudice:1998bp,Giudice:1998xp,Randall:1998uk,Pomarol:1999ie}. 
Despite the necessary presence of a flavor structure that 
suppresses Supersymmetric contributions to experimental observables, next-generation flavor 
experiments will be sensitive to TeV-scale SUSY, as the mechanisms suppressing 
flavor and CP violation are not completely effective at the level of expected experimental sensitivity.  
In addition, if Supersymmetry is discovered via direct production at the LHC, then 
measurements in the flavor sector will be essential to determine the flavor structure of the 
underlying theory.  Conversely, flavor experiments may provide a discovery path, having sensitivity 
to TeV-scale SUSY scenarios that are difficult to observe at the LHC.  It is thus imperative to 
perform a comprehensive analysis to examine the sensitivity of future flavor experiments to 
terascale Supersymmetry, and compare that to expectations for Run II at the LHC.

In order to be inclusive and avoid prejudice on the modelling of physics at the high-scale, we 
examine terascale SUSY in a general framework that captures the phenomenology of a variety of 
SUSY-breaking models, as well as that of SUSY-breaking mechanisms which remain to be discovered.  
We thus consider the phenomenological Minimal Supersymmetric Standard Model 
(pMSSM)~\cite{Djouadi:1998di,Berger:2008cq}, a subspace of the MSSM designed to examine regions of
parameter space that have been unexplored in studies of more simplified models.
The pMSSM is constructed from the general 
R-parity conserving MSSM by imposing the following set of data-driven assumptions: ($i$) No new 
sources of CP violation, ($ii$) Minimal Flavor Violation at the electroweak scale so that flavor 
violation is proportional to the CKM mixing matrix elements, ($iii$) degenerate 
1\textsuperscript{st} and 2\textsuperscript{nd} generation sfermion masses, and ($iv$) negligible 
Yukawa couplings and $A$-terms for the first two generations.  This results in a 19-dimensional 
parameter space, containing ten scalar masses, three gaugino masses, $M_{1,2,3}$, the three third 
generation $A_{t,b,\tau}$-terms, and three parameters related to the SUSY Higgs potential 
$\mu$, $M_A$ and $\tan \beta$. 

In order to examine the pMSSM contributions to CP-violating 
observables in this study, we augment the parameter space to include non-zero CP-violating phases.  
Including non-vanishing CP phases is particularly interesting because of their possible role in 
baryogenesis, as well as their potential to slightly increase the predicted value of the light Higgs mass within SUSY
\cite{Carena:2001fw}.  In particular, 
CP-violating phases can relieve some of the tension between the requirement of heavy stops and/or 
large stop mixing needed to generate the observed Higgs mass, versus the light stops and small mixing 
preferred by naturalness.  Specifically, we explore the effect of including 
all six CP-violating phases that are consistent with the pMSSM assumption of minimal flavor violation: 
$\phi_1 \equiv \text{arg}(M_1)$, $\phi_2 \equiv \text{arg}(M_2)$, $\phi_\mu \equiv
\text{arg}(\mu)$, $\phi_t \equiv \text{arg}(A_t)$,
$\phi_b \equiv \text{arg}(A_b)$, and $\phi_\tau \equiv
\text{arg}(A_\tau)$. 
We then study the effects of these non-zero phases on a variety of CP-conserving and CP-violating 
processes. We find that flavor observables place important constraints on the allowed soft SUSY 
breaking parameters, despite the assumptions ($ii$)-($iv$) listed above.  Interestingly, in some 
rare processes, such as BR($B_s\to \mu^+ \mu^-$), we find that the inclusion of CP-phases can have 
the effect of bringing the SUSY contributions more in line with SM expectations, and hence in 
better agreement with experiment. We will see that flavor constraints on the CP-extended pMSSM 
provide an additional handle that could allow SUSY to be discovered by the next generation of  
low-energy experiments, even if the sparticle spectrum is out of kinematic reach at the LHC. Such low-energy 
experiments are thus seen as complementary to the direct LHC SUSY searches in a manner similar to 
both the direct and indirect searches for Dark Matter \cite{Cahill-Rowley:2014boa}. This work 
expands upon the Snowmass study presented in \cite{Berger:2013zca}.

\section{Analysis Procedure}
\label{sec:procedure}

For this study, we extend the CP-conserving pMSSM model sample produced in 
\cite{CahillRowley:2012cb} which contains a neutralino lightest supersymmetric particle (LSP). This 
sample corresponds to a set of models (\ie, points in the parameter space), generated by a random 
scan employing flat priors over the 19 pMSSM parameters.  The scan ranges, shown in 
Table~\ref{ScanRanges}, were selected to enable phenomenological studies at the 14 TeV LHC.  These 
models were subjected to a global set of collider, flavor, precision electroweak, dark matter, and 
theoretical constraints.  We note that the WMAP/Planck measurement of the dark matter relic density 
was only employed as an upper bound, so that the LSP need not saturate this value, allowing for the 
possibility of multi-component dark matter.  This procedure generated $\sim 225$k pMSSM models that can 
be adopted for further studies.  The signatures of this model sample at the 7, 8, and 14 TeV LHC 
were recently examined in \cite{Cahill-Rowley:2014twa} using a fast Monte Carlo simulation of the 
ATLAS SUSY analysis suite. In that work, the expected ability of ATLAS to observe each model for 
each center-of-mass energy was determined\footnote{We note that our projections for the sensitivity 
of the 14 TeV LHC to the pMSSM models under consideration include only the search channels expected to be 
the most powerful.}. In particular, it was found that models with light squarks and gluinos remain 
viable after the LHC Run I at 7 and 8 TeV.  These collider results will enable the comparison of the
discovery reach between direct searches at the 14 TeV LHC and indirect effects in future low-energy 
precision measurements.

To extend this previous study to include CP-violating phases, 
we choose a random subset of 1000 models from this large pMSSM model sample  
which survive the 7 and 8 TeV LHC searches, ensuring that each 
model is in agreement with the observed Higgs mass within 3 GeV, corresponding to the theoretical 
error on the prediction of the Higgs mass in Supersymmetry\cite{sven}.  These models have the following 
characteristics: half of these models (\ie, 
500 models) are predicted to be detectable at the 14 TeV LHC with 300 fb$^{-1}$, while the other 
half are expected to remain viable (\ie, unobserved) even after 3000 fb$^{-1}$ at 14 TeV as 
described in \cite{Cahill-Rowley:2014twa}.  Since the CP-violating phases have minimal impact on 
the observability of models at the LHC, we choose to consider the effect of incorporating 
CP-violating phases to pMSSM models for which the LHC phenomenology has already been studied.

We now extend the 19-dimensional parameter space of each of these 1000 pMSSM models to include the phases 
present in the MSSM that are consistent with the pMSSM flavor structure. As discussed above, generically, six 
combinations of MSSM parameters can take on physical CP-violating phases; these combinations can be 
chosen to be $M_1 \mu B_\mu^*$, $M_2 \mu B_\mu^*$, $M_3 \mu B_\mu^*$, $A_t M_3^*$, $A_b M_3^*$, and 
$A_\tau M_3^*$.  We choose to work in a basis where $M_3$ and $B_\mu$ are real, so that the 
parameters with physical phases are $M_1$, $M_2$, $\mu$, $A_t$, $A_b$ and $A_\tau$. For each of 
the 1000 models we perform a random scan over these six physical phases, generating an 
additional 1000 models in each case, resulting in a total of $10^6$ models with CP-violating phases. 
The scan over the phases incorporates a 
log-uniform distribution over the 
range $10^{-6} \pi/2$ to $\pi/2$, with a random sign, as summarized in table~\ref{ScanRanges}.  This 
choice facilitates the study of a wider range of possible phases than that obtainable with a simple 
uniform scan distribution.  The overall sign of the parameters corresponding to these phases is 
fixed in the pMSSM.  The choice of a $\pi/2$ upper bound on the magnitude of the phase preserves 
the sign of the real part of the corresponding parameters. To restate: the result of this scan 
yields a total of 1 million models with non-zero CP-violating phases available for study.

\begin{table}
\centering
\begin{tabular}{|c|c|} \hline\hline
$m_{\tilde L(e)_{1/2,3}}$ & $100 \gev$ - $4 \tev$ \\ 
$m_{\tilde Q(u,d)_{1/2}}$ & $400 \gev$ - $4 \tev$ \\ 
$m_{\tilde Q(u,d)_{3}}$ &  $200 \gev$ - $4 \tev$ \\
$|M_1|$ & $50 \gev$ - $4 \tev$ \\
$|M_2|$ & $100 \gev$ - $4 \tev$ \\
$|\mu|$ & $100 \gev$ - $4 \tev$ \\ 
$M_3$ & $400 \gev$ - $4 \tev$ \\ 
$|A_{t,b,\tau}|$ & $0$ \gev - $4 \tev$ \\ 
$M_A$ & $100 \gev$ - $4 \tev$ \\ 
$\tan \beta$ & $1$ - $60$ \\
\hline
\vspace{3pt}
$\phi_1$ & $(10^{-6}$ - $1) \frac{\pi}{2}$ \\
$\phi_2$ & '' \\
$\phi_\mu$ & '' \\
$\phi_t$ & '' \\
$\phi_b$ & '' \\
$\phi_{\tau}$ & '' \\

\hline\hline
\end{tabular}
\caption{Scan ranges for the 25 parameters of the phase-extended pMSSM with a neutralino LSP. 
Mass parameters and $\tan \beta$ are scanned with flat priors, while the phases are scanned 
with log priors.}
\label{ScanRanges}
\end{table}
In what follows, we will refer to several different categories of models within our $10^6$ 
model sample of 
the phase-extended pMSSM. For clarity and ease of reference, the various model sub-categories 
considered in this work are summarized in Table \ref{ModelSets}. The original 1000 pMSSM models 
without phases are denoted as set A, divided into the half that is observable at the 14 TeV LHC 
(A1), and the half that is expected to evade the 14 TeV SUSY searches (A2).  The full set of $10^6$  
models in the phase-extended pMSSM is designated as set B. B1(B2) refers to the 
phase-extended models that are derived from the zero-phase pMSSM A1(A2) models that are observable 
(not observable) at the 14 TeV LHC, as detailed above.

\begin{table}
\centering
\begin{tabular}{|c|c|l|} \hline
Set name & \# of models & Description \\
\hline\hline
A & 1000 & Full pMSSM model set without phases \\
A1 & 500 & pMSSM models to which LHC-14 will be sensitive \\
A2 & 500 & pMSSM models which evade LHC-14 constraints \\
\hline
B & $10^6$ & Full model set A extended to include random phases\\
B1 & $5 \times 10^5$ & A1 with random phases \\
B2 & $5 \times 10^5$ & A2 with random phases\\
\hline
C & 155,474 & Number from set B allowed by current flavor \& CP constraints\\
C1 & 75,216 & Number from B1 allowed by current flavor \& CP 
constraints \\ 
C2 & 80,258 & Number from B2 allowed by current flavor \& CP constraints \\
\hline
D & 3708 & Number from set B allowed after future flavor \& CP null results\\
D1 & 1714 & Number from B1 allowed after future flavor \& CP null results\\
D2 & 1994 & Number from B2 allowed after future flavor \& CP null results\\
\hline
\end{tabular}
\caption{List of model categories referred to in the text along with the
  number of models in the category and a brief description.  For further
  details, see the text.}\label{ModelSets}
\end{table}
In this study, we employ the \verb+SUSY_FLAVOR v2.10+ code \cite{Crivellin:2012jv,Buras:2004qb,Dedes:2008iw} 
to calculate a comprehensive 
set of low-energy observables for the $10^6$ models in set B.  We performed an extensive number 
of analyses to verify the consistency of the output of the \verb+SUSY_FLAVOR v2.10+ code. The full 
list of processes we study is given in Table \ref{tab:obs}, along with the SM prediction for each 
observable as calculated by \verb+SUSY_FLAVOR+, the current experimental result, and the expected 
future experimental plus theory uncertainties or bounds as applicable.  We note that in many cases, 
the expected future measurements will provide a vast improvement over current sensitivity. The 
values of the 
input parameters, and their sources, that are required for the \verb+SUSY_FLAVOR+ computations are 
listed in Table \ref{tab:input}. Where they overlap, we took the input parameters to be identical 
to those used to generate the corresponding pMSSM models. The remaining parameters were chosen 
according to recent measurements, global data fits, or lattice calculations.

The observables listed in Table \ref{tab:obs} are the most constraining flavor and CP-violating processes 
computed by \verb+SUSY_FLAVOR+.  While the focus of this work is on CP-violating observables, the addition of 
CP-violating phases can also modify the pMSSM predictions for flavor changing, CP-conserving 
transitions due to the presence of $\phi$-dependent terms in the overall rates.  
Furthermore, only some of the low-energy constraints in Table \ref{tab:obs} have been applied 
during the generation of the pMSSM models in previous studies~\cite{CahillRowley:2012cb}.  The 
constraints from these processes offer a more complete picture of the current 
low-energy restrictions on the pMSSM.

Beginning with the original set B of $10^6$ models, we first determine the subset of these models that satisfy 
all of the existing flavor and CP constraints: We call this set C. Similarly, beginning with the B1(B2) 
set we derive those denoted as C1(C2).  Finally, we also consider the subset of B models that are expected to
survive the {\it future} flavor and CP constraints according to our study below, and refer 
to that subset as set D with analogously defined D1 and D2 subsets.

\begin{table}[!tb]
\begin{center}
\begin{tabular}{|l|p{2.6cm}r|p{2.4cm}r|lr|} 
\hline
\multicolumn{1}{|c|}{\multirow{2}{*}{Observable}} 
& \multicolumn{2}{c|}{\multirow{2}{*}{SM}} 
& \multicolumn{2}{c|}{\multirow{2}{*}{Experiment$^*$}} 
& \multicolumn{2}{c|}{Future Thy. $\oplus$ Exp.}\\
&  &  & &  & \multicolumn{2}{p{3.2cm}|}{bound/uncertainty}  \\
\hline
\hline
$|d_e|$ [e.cm]  & $\lesssim {\cal O}(10^{-40})$ 
& \cite{Pospelov:1991zt,Booth:1993af} 
&$<8.7\times10^{-29}$ & \cite{Aad:2012tx} & $ \lesssim 10^{-30}$  
& \cite{Vutha:2009ux,Wundt:2012di,Hewett:2012ns} \\
\hline
$|d_\mu|$ [e.cm] & $\lesssim {\cal O}(10^{-38})$ 
& \cite{BowserChao:1997bb} 
&$ < 1.6 \times 10^{-19}$& \cite{Bennett:2008dy} & $ \lesssim 10^{-24}$ 
& \cite{Farley:2003wt,Hewett:2012ns} \\
\hline
$|d_n|$ [e.cm] &  $\lesssim {\cal O}(10^{-31})$  
& \cite{Shabalin:1982sg,McKellar:1987tf} 
& $ < 2.9 \times 10^{-26}$ & \cite{Baker:2006ts} 
& $ \lesssim 5 \times 10^{-28}$ & 
\cite{Altarev:2009zz} \\
\hline
\m{$a_e$ [$10^{-12}$]} & $ 1159652182.79   $ & \m{\cite{Aoyama:2007mn}} 
& $1159652180.73$ & \m{\cite{Hanneke:2008tm}} &
\m{$\phantom{aaaaaa}--$} & 
\\
& $\phantom{aaaaa}\pm7.71$ &  & $\phantom{aaaaaaa}\pm28$ &  &  &   \\
\hline
\m{$a_\mu$ [$10^{-11}$]} & $116591802 \pm 49$ & \m{\cite{Blum:2013xva}} 
& \m{$116592089 \pm 63$} & \m{\cite{Bennett:2006fi}} & \m{$  \pm 12 $} 
& \m{\cite{Hewett:2012ns}} \\
& $116591828 \pm 50$ &  &  &  &  &   \\
\hline
\m{Br$(K_L \to \pi^0 \, \nu \, \bar{\nu})$} & $3.04 \pm 15.7\%$ 
&  \m{\cite{Mescia:2007kn}} & \m{$ < 2.6 \times 10^{-8}$} 
& \m{\cite{Ahn:2009gb}} & $ \pm 6.0\% \pm 5.1\% = $ & \m{\cite{KOTO,Hewett:2012ns}} \\
&  $\phantom{aaaa}\times 10^{-11}$ &  & & & $7.9\% $ & \\
\hline
Br$(K^+ \to \pi^+ \, \nu \, \bar{\nu})$ & \m{$9.2 \pm 8.2\%$} 
& \m{\cite{Mescia:2007kn}} & \m{$17.3^{+11.5}_{-10.5}$} 
& \m{\cite{Artamonov:2008qb}} & $ \pm 5.4\% \pm 2.2\% = $
& \m{\cite{E.T.WorcesterfortheORKA:2013cya,Hewett:2012ns}} \\
\phantom{aaaaaaaa} $\times 10^{11}$ &  & &  &  & $5.8\%$ &   \\
\hline
Br$(B_d \to X_s \, \gamma)^\dagger$ & \multirow{2}{*}{$3.17 \pm 7.3\%$} 
& \multirow{2}{*}{\cite{Misiak:2006zs}}& \multirow{2}{*}{$3.43 \pm 6.7\%$} 
& \multirow{2}{*}{\cite{Amhis:2012bh}} & $ \pm 6.7\% \pm
  4.0\% = $
& \m{\cite{belle2}} \\
\phantom{aaaaaaaa} $\times 10^{4}$ &  & &  &  & $7.8\%$ &  \\
\hline
Br($B_s \to \mu^+ \mu^-$) & \m{$3.74 \pm 4.1\%$} 
& \m{\cite{Bobeth:2013uxa}}
& \m{$2.9 \pm 0.7$} & 
\m{\cite{CMS-PAS-BPH-13-007}} & $  \pm 3.2\% \pm 8.6\% =$
& \m{\cite{CERN-LHCC-2011-001}} \\
\phantom{aaaaaaaa} $\times 10^{9}$ &  & &  &  & $ 9.2\%$ &   \\
\hline
Br($B_d \to \mu^+ \mu^-$) & \m{$1.21 \pm 6.1\%$} 
& \m{\cite{Bobeth:2013uxa}}
& \m{$3.6^{+1.6}_{-1.4}$} & \m{\cite{CMS-PAS-BPH-13-007}} 
& $  \pm 3.9\% \pm 36.0\% =$ & \m{\cite{CERN-LHCC-2011-001}} \\
\phantom{aaaaaaaa} $\times 10^{10}$ &  & &  &  & $ 36.2\%$ &   \\
\hline
Br($B_u \to \tau \nu_\tau$) & \m{$0.779 \pm 8.6\%$}  & \m{\cite{Bona:2009cj}}
& \multirow{2}{*}{$1.14 \pm 0.22$} & \multirow{2}{*}{\cite{Amhis:2012bh}} 
& $  \pm 6.0\% \pm 6.3\% =$  & \multirow{2}{*}{\cite{belle2}} \\
\phantom{aaaaaaaa} $\times 10^{4}$ & 
& &   &   & $8.7\%$  &     \\
\hline
\m{$\Delta M_{B_d}$[${\rm ps}^{-1}$]} & \m{$0.545 \pm 16.8\%$}
& \m{\cite{Lenz:2012mb}}
& \m{$ 0.507 \pm 0.005$}
& \m{\cite{Amhis:2012bh}} & $ \pm 3.7\% \pm 0.9\% = $ & \\
& & & & & $3.8\%$ & \\
\hline
\m{$\Delta M_{B_s}$[${\rm ps}^{-1}$]} & \m{$17.70 \pm 15.0\%$}
& \m{\cite{Lenz:2012mb}}
& \m{{\small $17.719\pm 0.043$}}
& \m{\cite{Amhis:2012bh}} & $ \pm 3.1\% \pm 0.2\% = $ & \\
& & & & & $3.1\%$ & \\
\hline
$\Delta M_K$[$10^{-3} \; {\rm ps}^{-1}$] & $4.824$ &  & $5.292 \pm 0.009$ 
& \cite{Beringer:1900zz} & $\phantom{aaaaaa}--$ & \\
\hline
\m{$\epsilon_K$[$10^{-3}$]} & \m{$2.319\pm 9.3\%$} & \m{\cite{utfit}} & \m{$2.228 \pm 0.011$} & 
\m{\cite{Beringer:1900zz}} 
& \m{$\phantom{aaaaaa}--$} & \\
& & & & & &   \\
\hline
\m{sin$(2\beta)$} & \m{$0.695 \pm 5.6\%$} & \m{\cite{utfit}}
& \m{$0.68 \pm 0.02$} 
& \m{\cite{Amhis:2012bh}} &   $\pm 2.1\% \pm 1.2 \% = $
& \m{\cite{CERN-LHCC-2011-001}} \\
& & & &  &  $2.4\%$ &  \\
\hline
\m{sin$(2\beta_s)$} & \m{$0.0375 \pm 4.0\%$} &  \m{\cite{utfit}}
& \m{$-0.04^{+0.13}_{-0.10}$} & \m{\cite{Amhis:2012bh}} 
& $ \pm 2.5\% \pm 15.8\% = $ & \m{\cite{CERN-LHCC-2011-001}} \\
& &&  &  & $\pm 16.0\%$ &   \\
\hline
\multicolumn{7}{l}{$^*$All upper bounds are at 90\% C.L., \,
$^\dagger E_\gamma > 1.6$ GeV in the $B$-meson rest frame. }
\end{tabular}
\caption{The complete set of observables studied in this work. All processes 
are computed using SUSY\_FLAVOR v2.10.\label{tab:obs}}
\end{center}
\end{table}

\begin{table}[!tb]
  \centering
  \begin{minipage}[t]{0.45\textwidth}
  \begin{tabular}[t]{c c}
    \hline
    Observable & Value \\
    \hline
    \hline
    \multicolumn{2}{l}{pMSSM input} \\
    \hline
    $\alpha^{-1}(m_Z)$ & $127.8568$~\cite{CahillRowley:2012cb} \\
    $\alpha_s(m_Z)$ & $0.1193$~\cite{Baak:2011ze} \\
    $m_Z$ & $91.1876~{\rm GeV}$~\cite{Beringer:1900zz} \\
    $m_W$ & $80.385~{\rm GeV}$~\cite{Beringer:1900zz} \\
    $m_b(m_b)$ & $4.16~{\rm GeV}$~\cite{Chetyrkin:2009fv} \\
    $m_t^{\rm pole}$ & $173.2~{\rm GeV}$~\cite{Lancaster:2011wr} \\
    \hline
    \hline
    \multicolumn{2}{l}{Quark masses and $D$-meson mass \cite{Beringer:1900zz}} \\
    \hline
    $m_u(2~{\rm GeV})$ & $2.3~{\rm MeV}$ \\
    $m_d(2~{\rm GeV})$ & $4.8~{\rm MeV}$ \\
    $m_s(2~{\rm GeV})$ & $95~{\rm MeV}$ \\
    $m_e$ & $0.510998928~{\rm MeV}$ \\
    $m_\mu$ & $105.659~{\rm MeV}$ \\
    $m_\tau$ & $1.777~{\rm GeV}$ \\
    $m_D$ & $1.8645~{\rm GeV}$ \\
    $\Delta m_D$ & $1.56 \times 10^{-14}~{\rm GeV}^{-1}$ \\
    \hline
    \hline
    \multicolumn{2}{l}{CKM \cite{Ciuchini:2000de}} \\
    \hline
    $\lambda$ & $0.22535$ \\
    $A$ & $0.822$ \\
    $\bar{\rho}$ & $0.127$ \\
    $\bar{\eta}$ & $0.353$ \\
    \hline
    \hline
    \multicolumn{2}{l}{Experimental inputs \cite{Buras:2013ooa}} \\
    \hline
    $m_c(m_c)$ & $1.279~{\rm GeV}$ \\
    $m_K$ & $497.614~{\rm MeV}$ \\
    $m_{B_d}$ & $5.2792~{\rm GeV}$ \\
    $m_{B_s}$ & $5.3668~{\rm GeV}$ \\
    $\tau_{B_d}$ & $1.519~{\rm ps}^{-1}$ \\
    $\tau_{B_s}$ & $1.516~{\rm ps}^{-1}$ \\
    $\Delta m_K$ & $3.483 \times 10^{-15}~{\rm GeV}^{-1}$ \\
    $\Delta m_{B_d}$ & $3.36 \times 10^{-13}~{\rm GeV}^{-1}$ \\
    $\Delta m_{B_s}$ & $1.164 \times 10^{-11}~{\rm GeV}^{-1}$ \\
    $\epsilon_K$ & $2.228 \times 10^{-3}$ \\
    \hline
    \hline
    \multicolumn{2}{l}{$n$ EDM \cite{Crivellin:2012jv}} \\
    \hline
    $\eta_e$ & $1.53$ \\
    $\eta_c$ & $3.4$ \\
    $\eta_g$ & $3.4$ \\
    $\Lambda_X$ & $1.18~{\rm GeV}$ \\
    \hline
    \hline
    \multicolumn{2}{l}{Lattice Averages} \\
    \hline
    $f_D$ & $200~{\rm MeV}$ \\
    \hline
  \end{tabular}
  \end{minipage}
  \begin{minipage}[t]{0.45\textwidth}
  \begin{tabular}[t]{c c}
    \hline
    Observable & Value \\
    \hline
    \hline
    \multicolumn{2}{l}{Basic lattice parameters \cite{Buras:2013ooa}} \\
    \hline
    $f_K$ & $156.1~{\rm MeV}$ \\
    $f_{B_d}$ & $190.5~{\rm MeV}$ \\
    $f_{B_s}$ & $227.7~{\rm MeV}$ \\
    $\eta_{cc}$ & $1.87$ \\
    $\eta_{ct}$ & $0.496$ \\
    $\eta_{tt}$ & $0.5765$ \\
    $\eta_b$ & $0.55$ \\
    $\hat{B}_K$ & $0.766$ \\
    $\hat{B}_{B_d}$ & $1.27$ \\
    $\hat{B}_{B_s}$ & $1.33$ \\
    $\kappa_0$ & $(2.31 \pm 0.01) \times 10^{-10}$ \\
    $\kappa_+$ & $(5.36 \pm 0.026) \times 10^{-11}$ \\
    $P_c$ & $(0.42 \pm 0.03) \times 10^{-10}$ \\
    \hline
    \hline
    \multicolumn{2}{l}{$K$- and $D$-meson $B$ \cite{Carrasco:2012dd}} \\
    \hline
    $B_K^{VLL}(2~{\rm GeV})$ & $0.52$ \\
    $B_K^{SLL1}(2~{\rm GeV})$ & $0.54$ \\
    $B_K^{SLL2}(2~{\rm GeV})$ & $0.27$ \\
    $B_K^{LR1}(2~{\rm GeV})$ & $0.63$ \\
    $B_K^{LR2}(2~{\rm GeV})$ & $0.82$ \\
    $B_D^{VLL}(2~{\rm GeV})$ & $0.78$ \\
    $B_D^{SLL1}(2~{\rm GeV})$ & $0.71$ \\
    $B_D^{SLL2}(2~{\rm GeV})$ & $0.45$ \\
    $B_D^{LR1}(2~{\rm GeV})$ & $1.17$ \\
    $B_D^{LR2}(2~{\rm GeV})$ & $0.94$ \\
    \hline
    \hline
    \multicolumn{2}{l}{$D$-meson RG-invariant $B$} \\
    \hline
    $\hat{B}_D$ & $1.17$ \\
    \hline
    \hline
    \multicolumn{2}{l}{$B$-meson $B$ \cite{Carrasco:2013zta}} \\
    \hline
    $B_{B_d}^{VLL}(m_b)$ & $0.85$ \\
    $B_{B_d}^{SLL1}(m_b)$ & $0.72$ \\
    $B_{B_d}^{SLL2}(m_b)$ & $0.61$ \\
    $B_{B_d}^{LR1}(m_b)$ & $1.47$ \\
    $B_{B_d}^{LR2}(m_b)$ & $0.95$ \\
    $B_{B_s}^{VLL}(m_b)$ & $0.86$ \\
    $B_{B_s}^{SLL1}(m_b)$ & $0.73$ \\
    $B_{B_s}^{SLL2}(m_b)$ & $0.62$ \\
    $B_{B_s}^{LR1}(m_b)$ & $1.57$ \\
    $B_{B_s}^{LR2}(m_b)$ & $0.93$ \\
    \hline
  \end{tabular}
\end{minipage}
  \cprotect\caption{Values of Standard Model, lattice, and observational quantities we employ with 
    \verb+SUSY_FLAVOR v2.10+.\label{tab:input}}
\end{table}

In the following analysis, we will mainly investigate model set C, containing the 155,474 models  
for which the predicted values of the observables in Table \ref{tab:obs} are below the current 90\% CL 
limits or within 2$\sigma$ of the observed values for measured 
quantities.\footnote{We do not apply the experimental constraint on $(g-2)_\mu$ at this stage 
as it lies 3.4$\sigma$ deviation from the SM prediction. In the following analysis, we will 
consider the possibility that future measurements observe either the SM prediction, or the current 
central value, with increased precision.} Some models, particularly those with large values of $\tan\beta$, can 
generate significant corrections to the CKM matrix elements, as well as Yukawa couplings \cite{Isidori:2010kg}.  In these cases, 
$\tan\beta$ enhancements compensate for loop factor suppressions and can lead to large loop 
contributions with diminished perturbative control without, \eg, resummation.  Thus we enforce
a cut of $40\%$ on the size of the maximal correction to the CKM matrix elements or 
the Yukawa couplings in model set C, to ensure that perturbative control of the flavor calculations is maintained. 
Keeping with our previous nomenclature, we define C1 as the 80,258 models obtained by applying the same LHC 
constraints as for B1 and, 
correspondingly, C2 as the 75,216 models obtained by applying the same LHC constraints as in B2.  The most 
constraining observables at this stage are 
$d_e$, ${\rm Br}(B_s \to \mu \mu)$, and ${\rm Br}(B \to s \gamma)$.

We will also consider the models denoted by D, D1, and D2 that are expected to remain viable after
future flavor experiments are performed \textit{assuming}  the central measured value agrees with
the SM prediction.  The D model sets are comprised of 3708, 1994, and 
1714 models, respectively, where D1 and D2 represent the sets corresponding to the LHC discovery criteria described
above.  In this study, we employ the anticipated future experimental uncertainties given in \cite{Hewett:2012ns}.

For several observables, the improvement in future sensitivity is
dominated by the expected reduction in the error associated with the theoretical calculation of the
value for the process.  To estimate the theoretical errors, we
include the uncertainties from both the input parameters and those due to lattice calculations and/or higher order
terms in the expansion employed in the calculation.  While the present
uncertainties on the observables studied in this work are well documented in the
literature, we estimated the projected future uncertainties independently.

Specifically, we estimate the future theoretical uncertainties on an observable $O$
as follows.  In general, the calculation of $O$ can depend on model
parameters $\alpha_i$ which are expected to be determined with
uncertainties $\sigma(\alpha_i)$ with updated experimental input or improved calculations.  The
most important parameters $\alpha_i$ for our analysis are the CKM
angle $|V_{ub}|$ and the CKM phase $\delta$, although we include all the
parameters listed in Table \ref{tab:input}.  The uncertainty in $O$
due to the corresponding error in $\alpha_i$ can be estimated as 
\begin{equation}
  \label{eq:2}
  \sigma_i(O) = \frac{\partial O}{\partial \alpha_i} \sigma(\alpha_i).
\end{equation}
The computation of $O$ can also have an uncertainty due to the
approximations used to derive the amplitude for $O$, which we denote by $\sigma_{\rm th}(O)$.  Then the total
uncertainty $\sigma(O)$ for an observable can be determined by adding
$\sigma_i(O)$ and $\sigma_{\rm th}(O)$ in quadrature as
\begin{equation}
  \label{eq:1}
  \sigma^2(O) = \sigma_{\rm th}^2(O) + \sum_i \sigma_i^2(O) \,.
\end{equation}
Strictly speaking, the theoretical uncertainty is model dependent, however we assume the SM
and leading expressions for $O$ in our estimations.  The uncertainties estimated in this manner
are more conservative than those quoted in other studies, 
but generally
agree well (see Table \ref{tab:obs}).  In order to
obtain projections for future uncertainties, we assume the
improvements in the model parameters as outlined in
\cite{Hewett:2012ns}, while taking $\sigma_{\rm th}(O)$ to remain unchanged. 
The resulting error projections are shown in the
rightmost column of Table \ref{tab:obs}. 

It is worth noting already at this point that the relative sizes of the two subsets C1 and C2, as well as 
D1 and D2, are comparable.  This observed similarity is an indicator of a high degree of 
complementarity between the direct searches at the LHC and low-energy probes of the MSSM.  
A strong constraint from the LHC only biases the 
low-energy bounds by roughly $\sim 10\%$.

\section{Numerical Results}
\label{sec:results}

We now characterize the model sets C and D, the pMSSM models with CP phases satisfying the current and expected
future constraints, respectively, from low-energy observables.

\begin{figure}[h]
\begin{center}
\begin{tabular}{cc}
\begin{overpic}[scale=1]{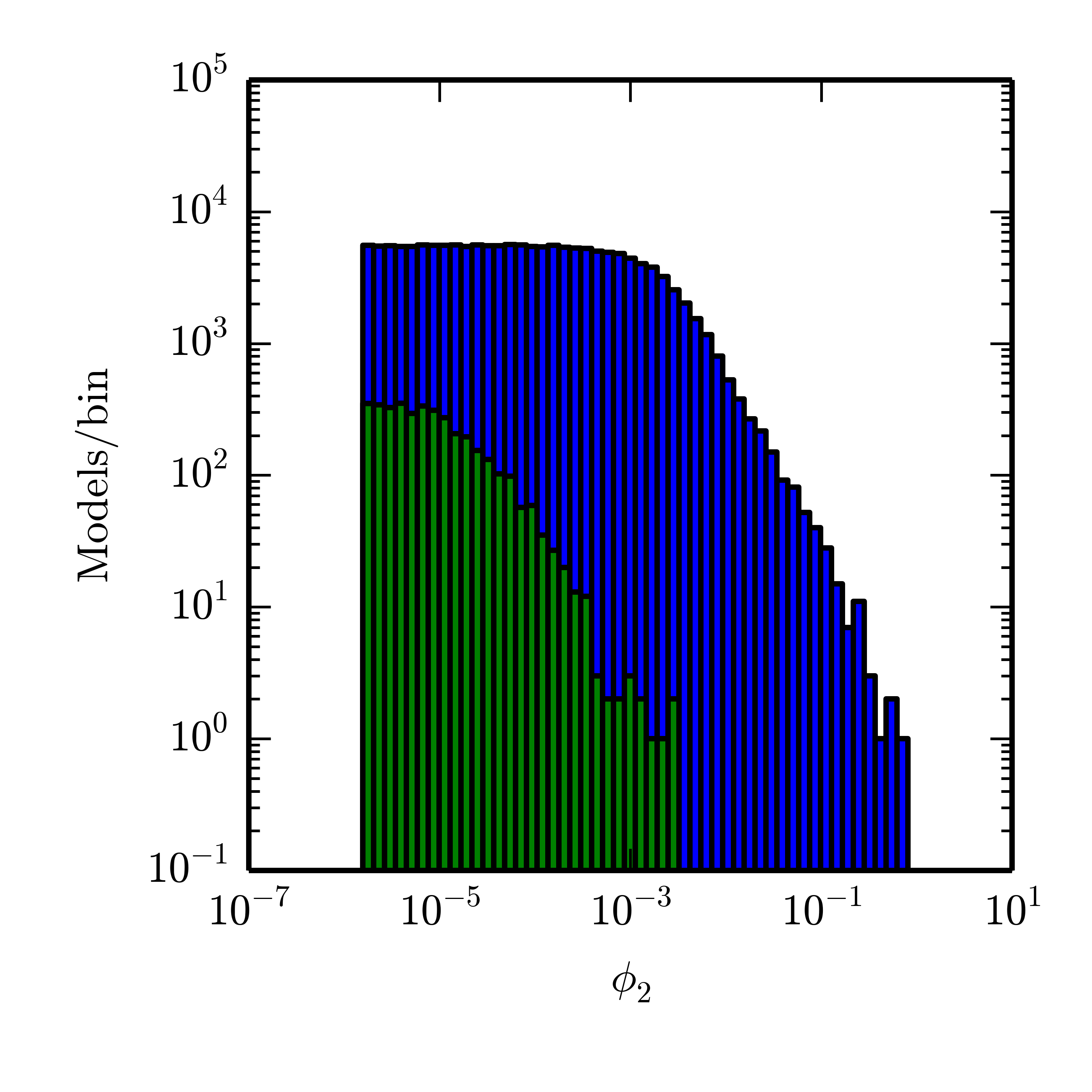} \put (40,83) {\Large (a)} \end{overpic} &
\begin{overpic}[scale=1]{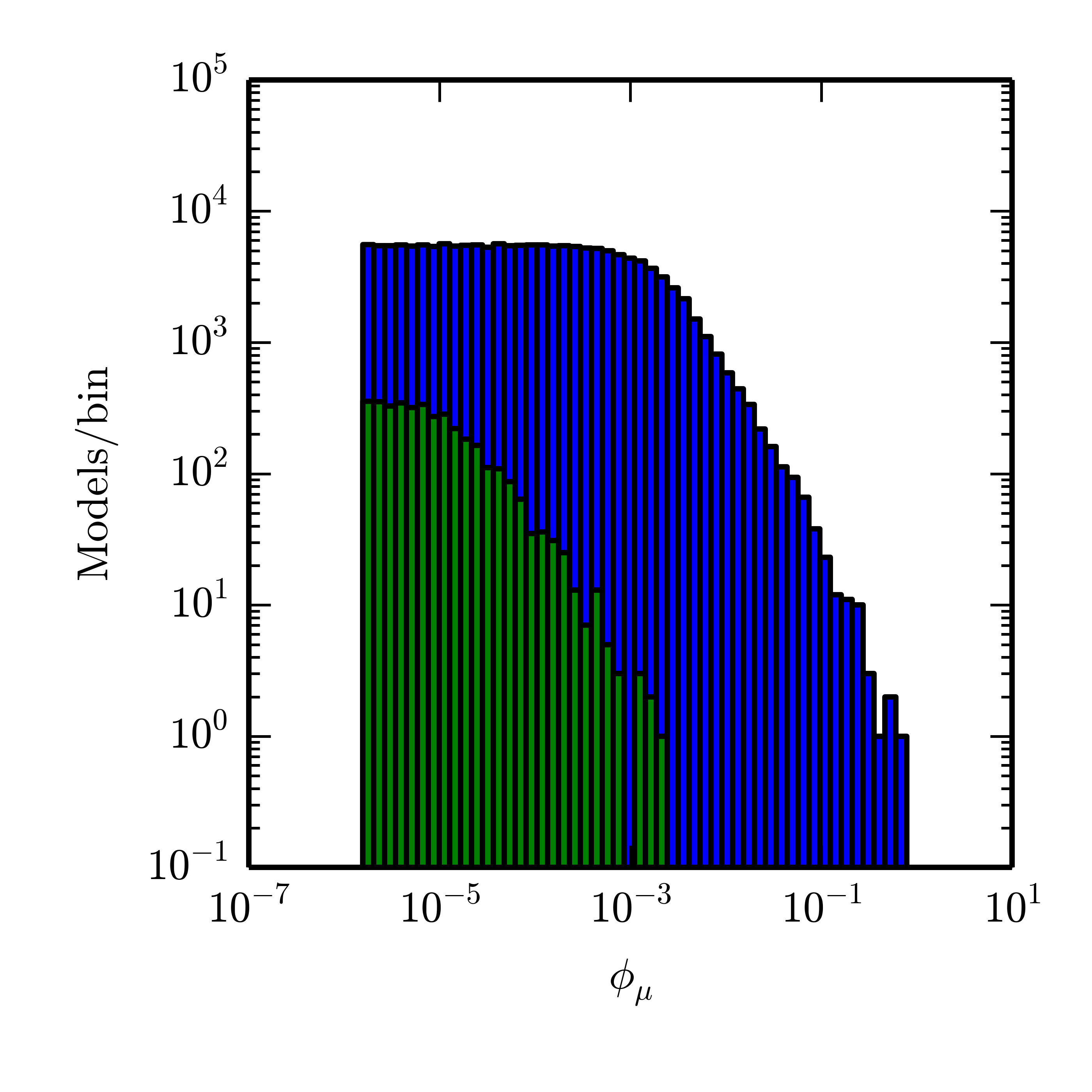} \put (40,83) {\Large (b)} \end{overpic} \\[-0.7cm]
\begin{overpic}[scale=1]{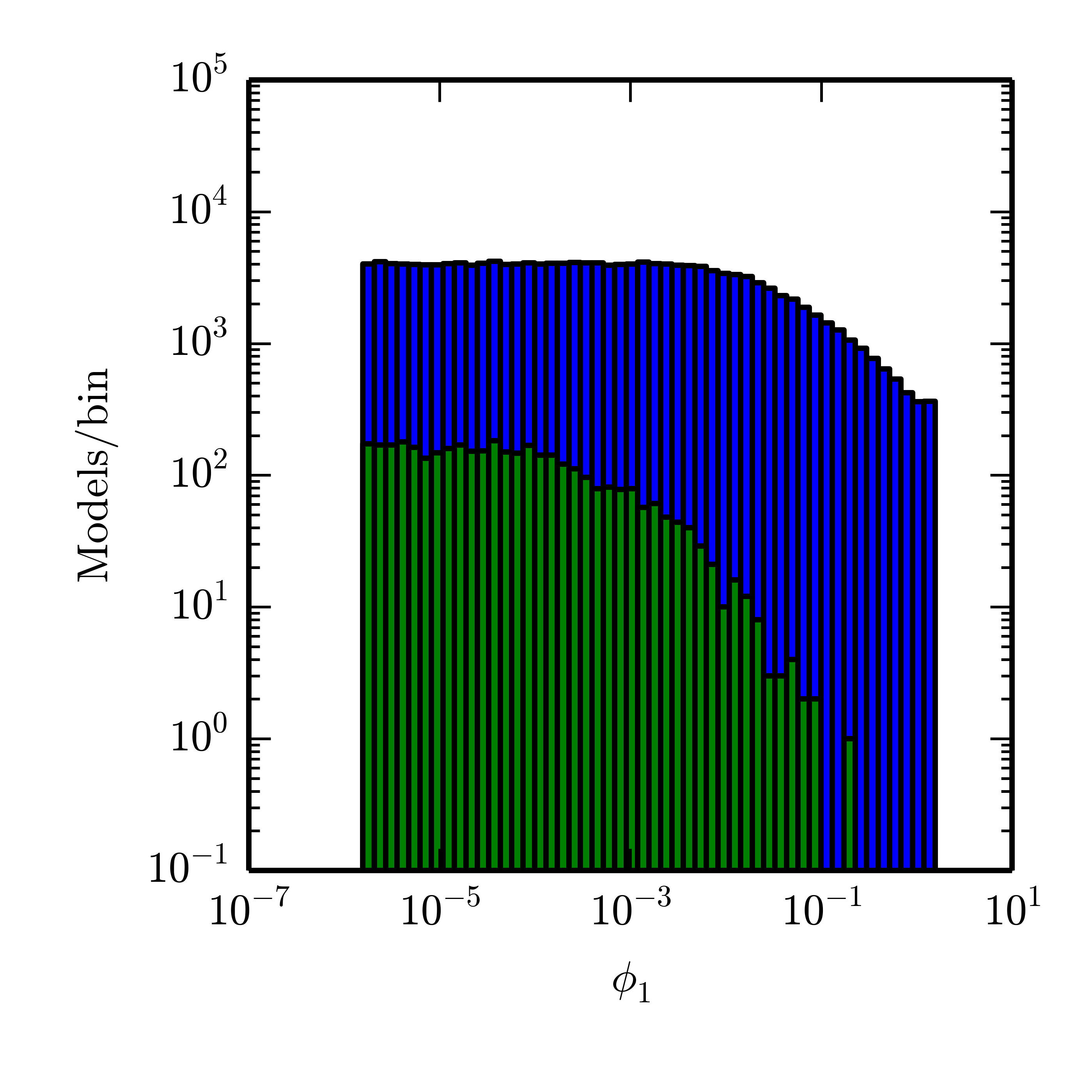} \put (40,83) {\Large (c)} \end{overpic} &
\begin{overpic}[scale=1]{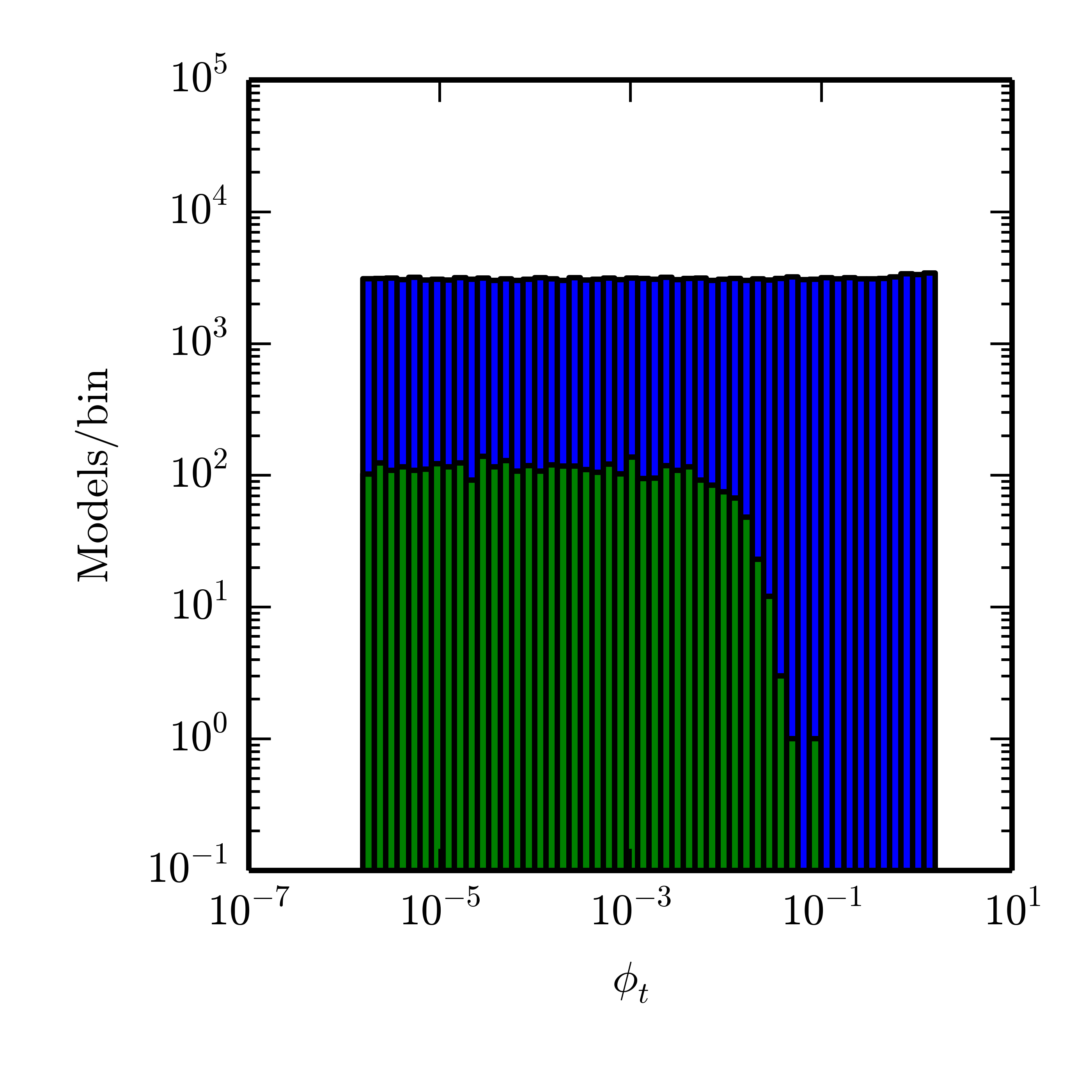} \put (40,83) {\Large (d)} \end{overpic} \\[-0.7cm]
\begin{overpic}[scale=1]{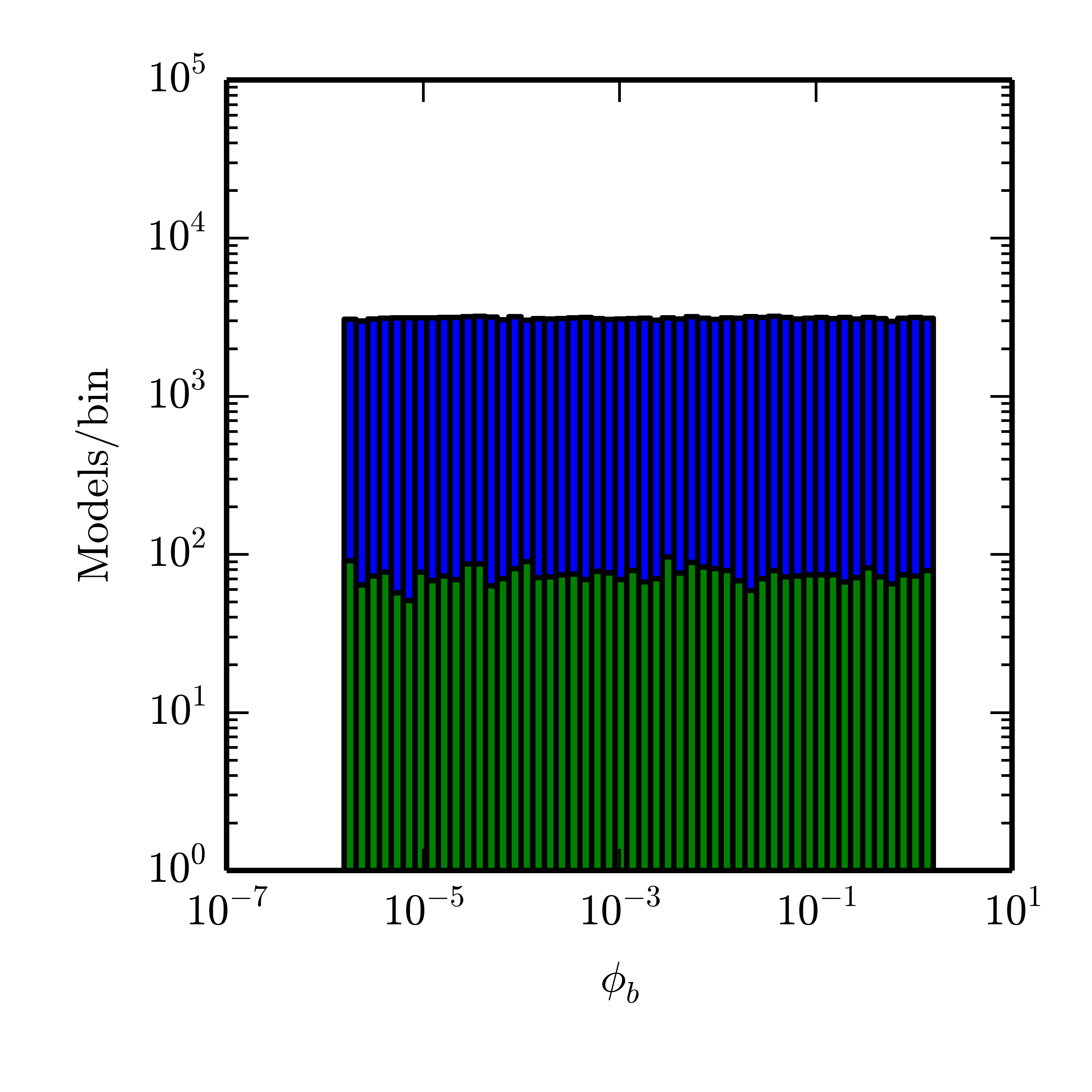} \put (40,85) {\Large (e)} \end{overpic} &
\begin{overpic}[scale=1]{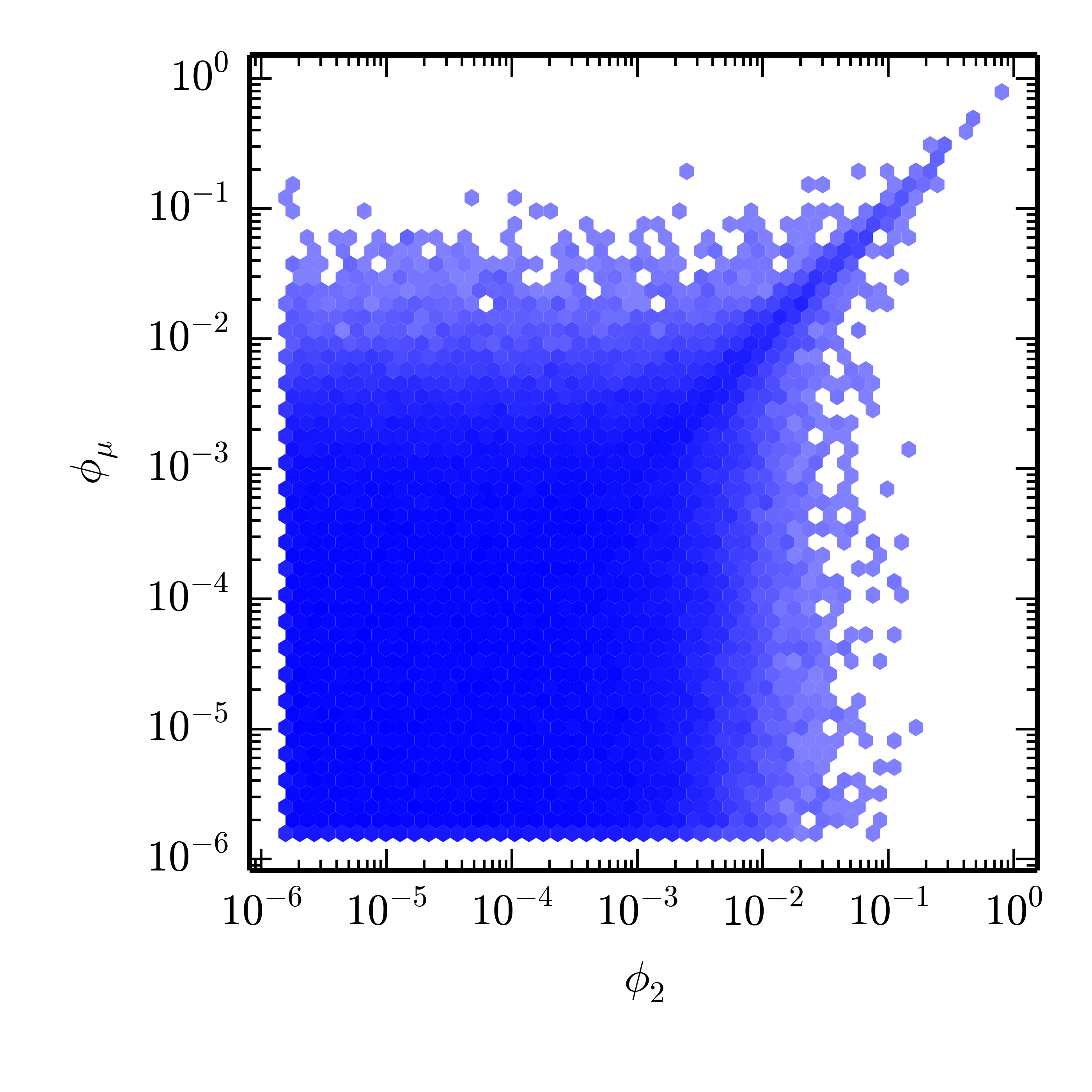} \put (40,85) {\Large (f)} \end{overpic}\\[-0.5cm]
\end{tabular}
\caption[]{(a)-(e): Distributions of the CP-violating phases $\phi_{1,2,\mu,t,b}$ added to the pMSSM
  in model sets C (blue) and D (green). (f): Model densities of values for the phases
  $\phi_2$ and $\phi_\mu$ in model set C. The shading ranges from light blue to
  dark blue with the darkness being logarithmic in the number of models in the bin.
  White bins contain no models.  The model sets are defined in Table \ref{ModelSets}.
\label{fig:1}}
\end{center}
\end{figure}

We begin our survey with an overview of the values of the phases in the CP-violating pMSSM. 
The first five panels of Figure \ref{fig:1} contain histograms of the values of the 
CP phases $\phi_{1,2,\mu,t,b}$ after 
applying the current (blue histogram, model set C) and future (green histogram, model set D) constraints on the observables 
in Table~\ref{tab:obs}. In showing the expected future constraints, we once again emphasize our assumption that 
experimental measurements will obtain the SM predictions given in Table~\ref{tab:obs}. These figures 
demonstrate the high sensitivity of future flavor experiments to the phases 
$\phi_1$, $\phi_2$, $\phi_\mu$ and $\phi_t$.  Null results from future experiments would require $\phi_2$ 
and $\phi_\mu$ to be small, ${\cal O}(10^{-2})$ or less, while weaker limits of ${\cal O}(10^{-1})$ would 
be placed on $\phi_1$ and $\phi_t$. Interestingly, ${\cal O}(1)$ values of $\phi_b$ would remain essentially unconstrained. 
We note that $\phi_\tau$ (not shown) also remains unconstrained.

Figure \ref{fig:1} displays a two-dimensional density histogram of the phases $\phi_2$ 
and $\phi_\mu$ for the models that pass current constraints (model set C) and illustrates interesting 
correlations. Note that there is a region with $\phi_2 \approx \phi_\mu$ where ${\cal O}(1)$ values of both
phases are allowed.  This corresponds to the case 
where the most significant 
effective phase contributing to the EDMs is effectively zero, due to a cancellation among gaugino 
exchange diagrams~\cite{Ibrahim:1998je}. Due to this 
cancellation, models with large and nearly equal phases $\phi_2$ and $\phi_\mu$ survive the EDM bounds. 
These models are, however, tuned in the sense that they are delicately sensitive to the degree of 
cancellation between these phases.

\begin{figure}[h]
\begin{center}
\begin{tabular}{cc}
\begin{overpic}[scale=1]{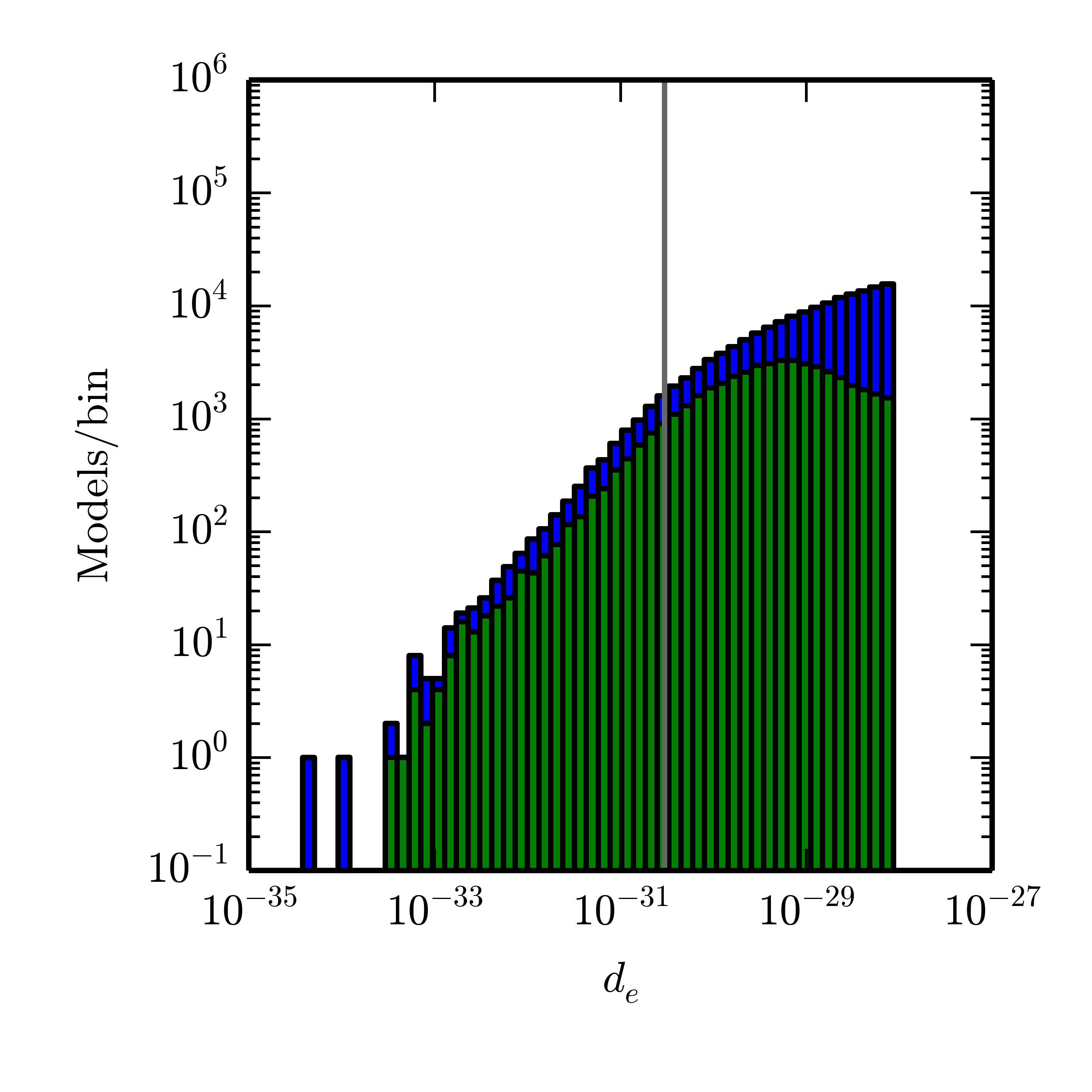} \put (40,80) {\Large (a)} \end{overpic} &
\begin{overpic}[scale=1]{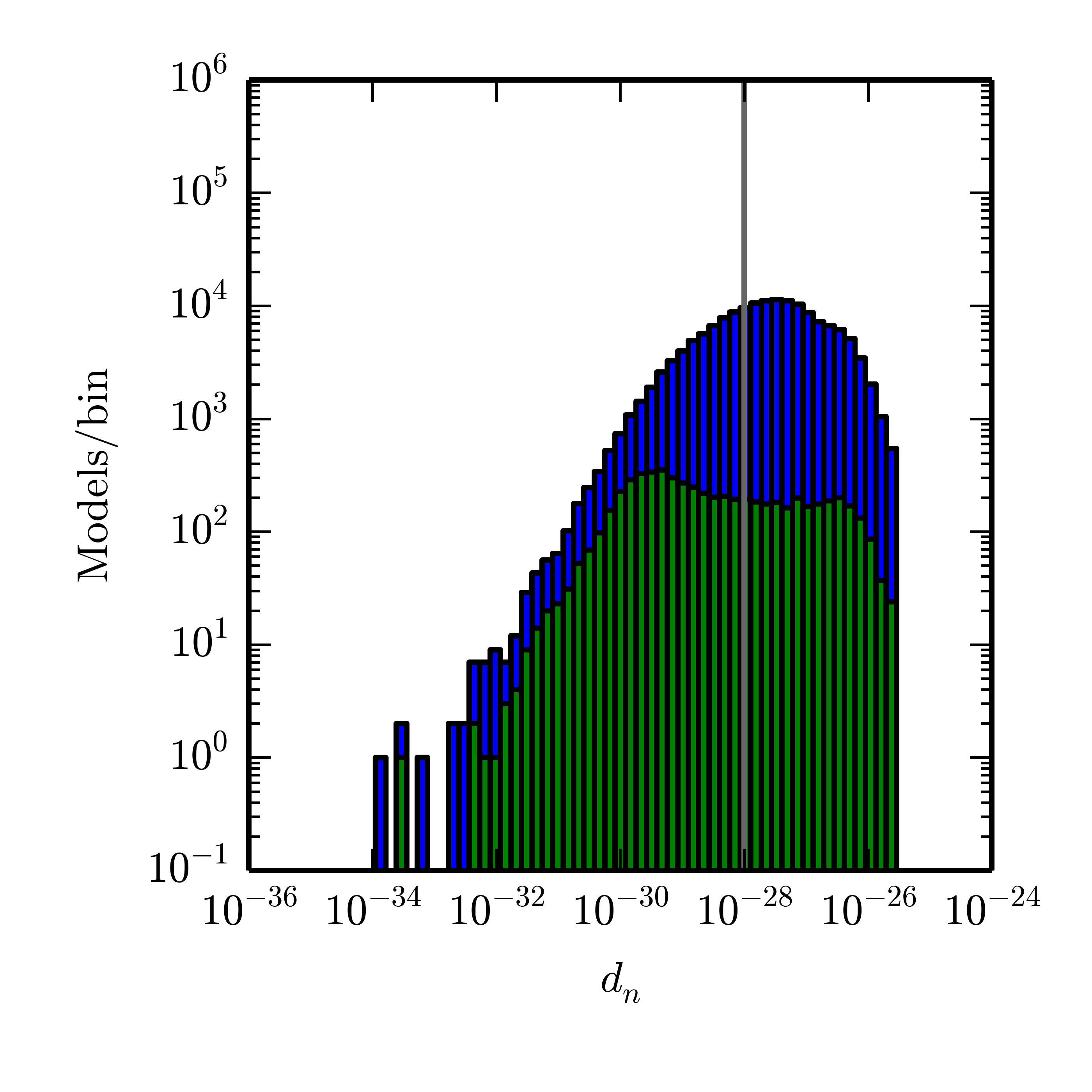} \put (40,80) {\Large (b)} \end{overpic} \\[-0.7cm]
\begin{overpic}[scale=1]{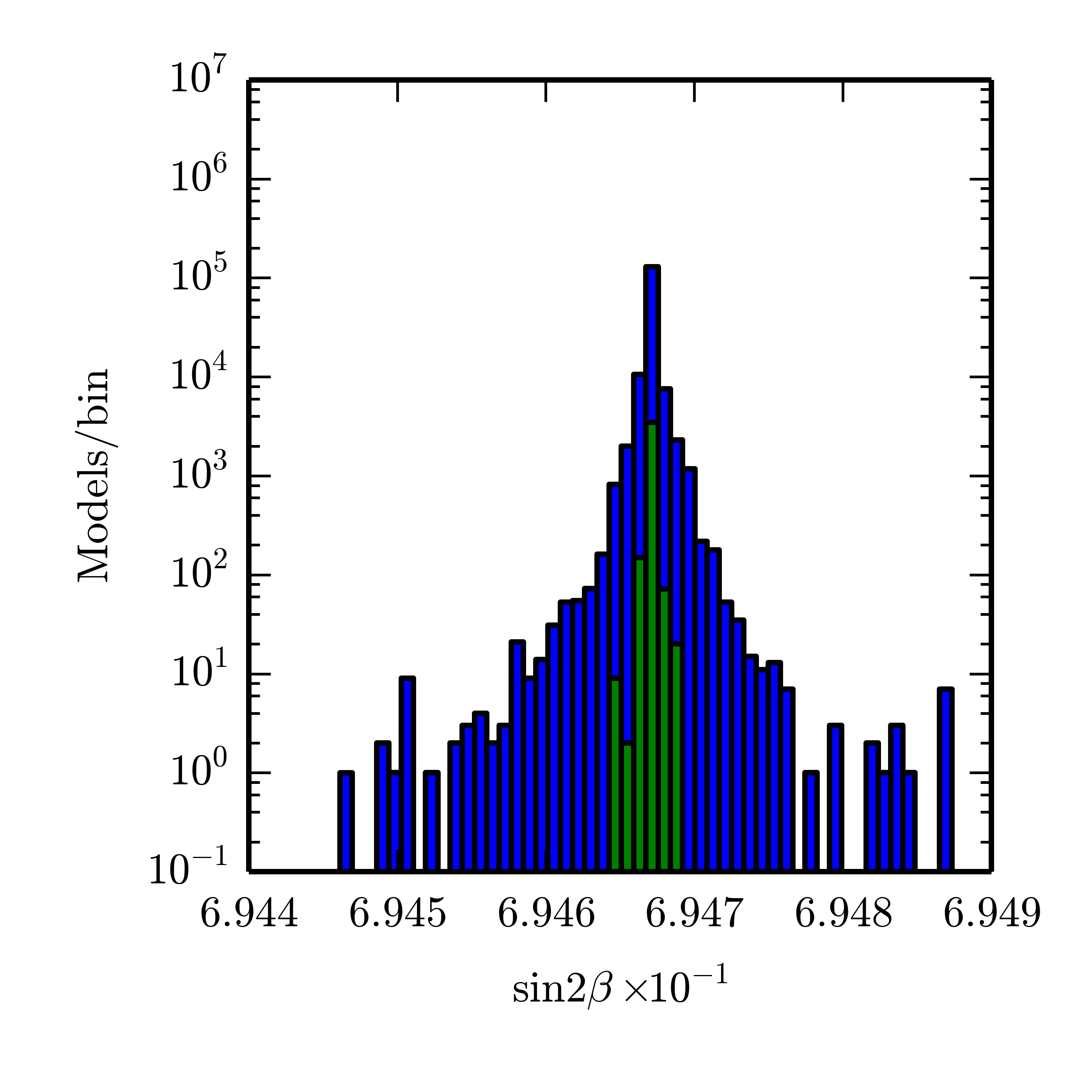} \put (40,80) {\Large (c)} \end{overpic} &
\begin{overpic}[scale=1]{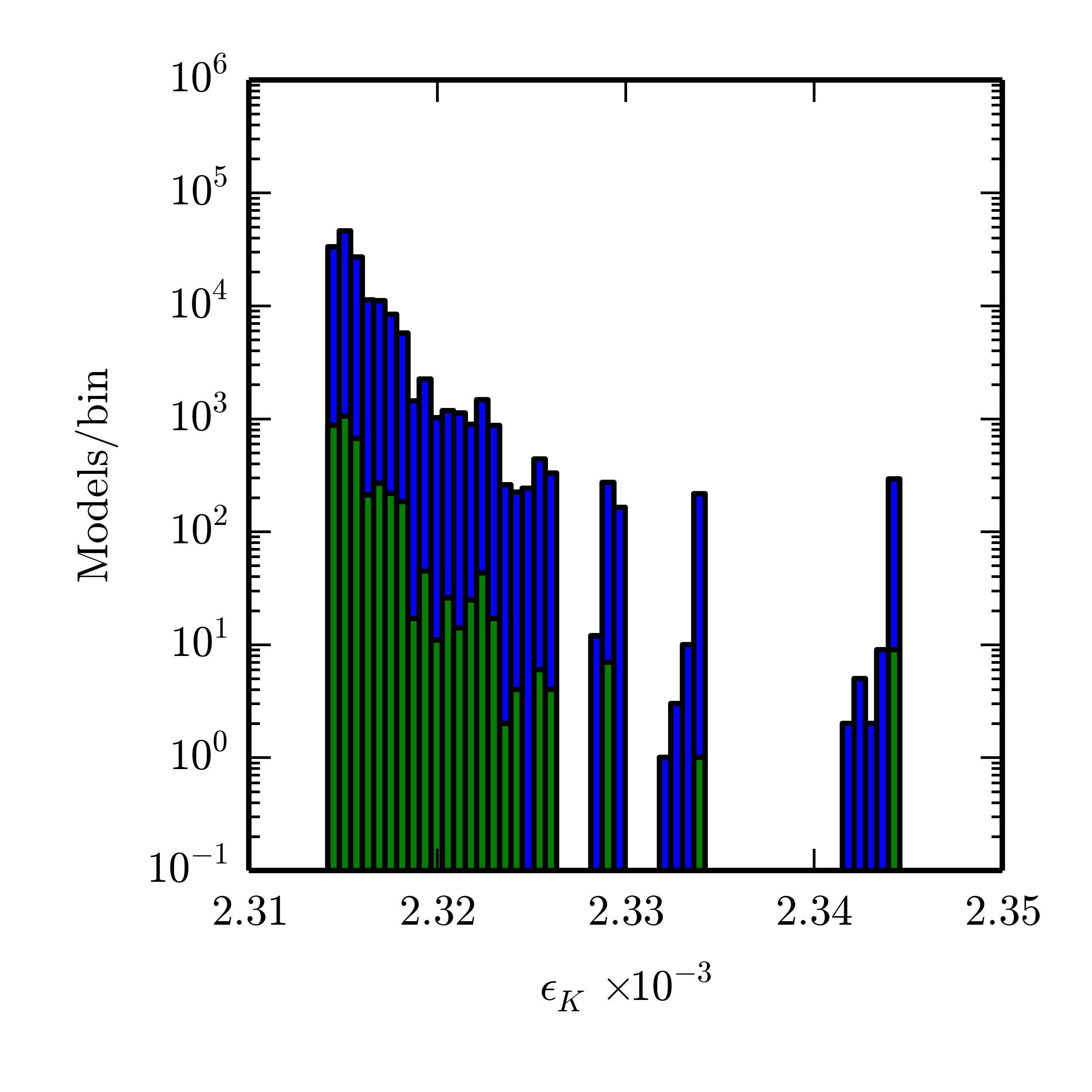} \put (40,80) {\Large (d)} \end{overpic} \\[-0.7cm]
\begin{overpic}[scale=1]{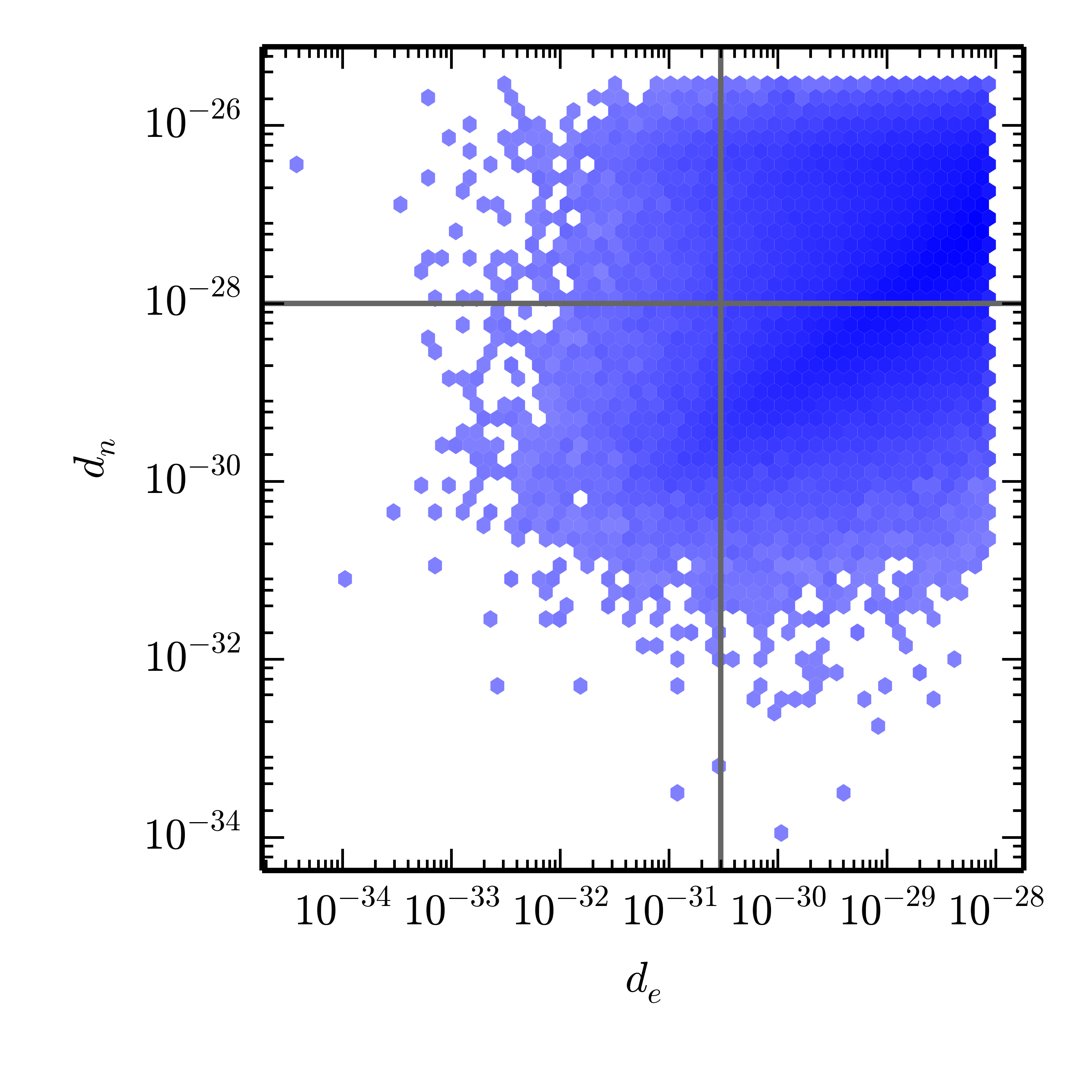} \put (30,25) {\Large (e)} \end{overpic} &
\begin{overpic}[scale=1]{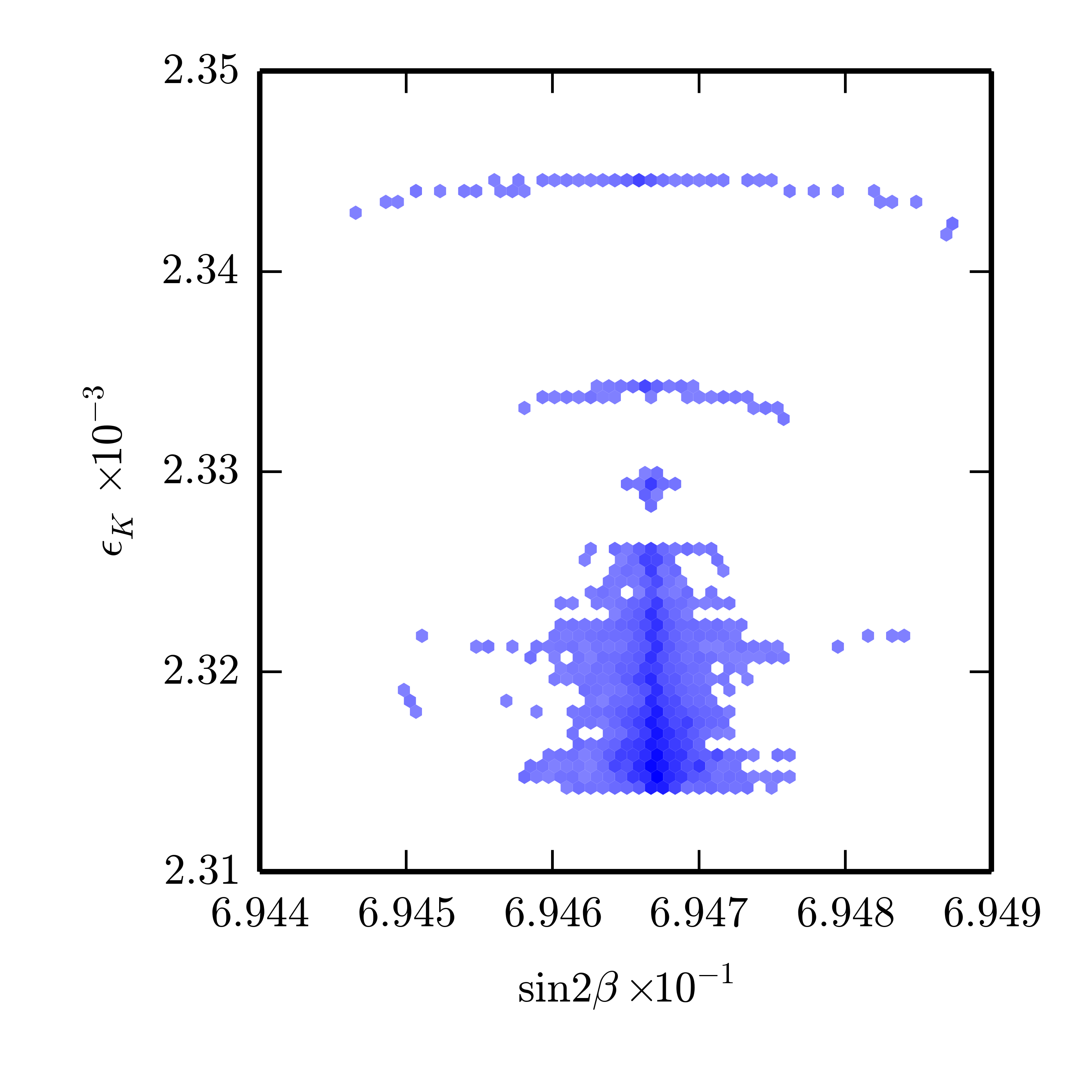} \put (30,25) {\Large (f)} \end{overpic}\\[-1cm]
\end{tabular}
\caption[]{(a)-(d): Expected constraints from future low-energy experiments on the
  pMSSM models with CP violation. In each panel, the
  top histogram shows models after applying current constraints (blue, model set C), while
  the bottom histogram (green) represents the remaining models after applying all future constraints except
  for the observable under study.  This illustrates the exclusive 
  ability of that observable to probe the model parameter space. Where possible, the anticipated future 
  limit on the observable is indicated by a vertical or horizontal line. (e)-(f): Model densities for 
  the observables in model set C. The shading is as in Figure\ref{fig:1}(f).
\label{fig:2}}
\end{center}
\end{figure}

The observables that are most sensitive to CP-violating effects are the electron and 
neutron EDMs, as well as $\sin2\beta$ and $\epsilon_K$. The upper four panels of Figure \ref{fig:2} show 
histograms of these quantities, demonstrating the potential of individual experiments to constrain 
these observables or detect new physics arising from SUSY. In each panel, the upper histogram corresponds 
to models that have passed all the current 
constraints (blue, model set C), while the lower histogram (green) is obtained by applying anticipated future constraints 
from all observables except for the one under study.  The expected future limit on the  EDMs 
is represented by the vertical lines. Here, we see the well-known 
feature that future EDM constraints will be very important in probing supersymmetric CP phases. We note, 
however,  that the expected future measurements for $\sin 2\beta$ and $\epsilon_K$ 
lie outside the ranges shown in the 
histograms, indicating that improved measurements of these quantities will not further 
constrain these models. We observe that there is a large correlation 
between $\sin2\beta$ and the EDMs, as the models passing future 
constraints (dominated by EDM limits) cluster around the SM expectations for $\sin 2\beta$. 
\clearpage
%

%
\clearpage
If future EDM experiments continue to 
obtain null results, a non-SM measurement of $\sin2\beta$ would thus pose an intriguing challenge to the 
CP-violating pMSSM. On the other hand, we see that $\epsilon_K$ is not particularly 
sensitive to the gaugino phases and is therefore poorly correlated with the other CP-violating 
observables. A relatively large deviation from the SM in $\epsilon_K$ 
would still be possible, even if future EDM searches are null.
The lower two panels of Figure \ref{fig:2} display two-dimensional histograms of these observables 
for models in agreement with the current experimental constraints (model set C).  The lines in figure 2e correspond to the 
anticipated future limits listed in Table 3.
These figures serve to further illustrate the 
correlations among CP-violating observables. We see that the electron and neutron EDMs are strongly 
aligned as expected, while there is no correlation between $\sin2\beta$ and 
$\epsilon_K$, due to the latter's lack of sensitivity to the gaugino phases.

\begin{figure}[]
\begin{center}
\begin{tabular}{cc}
\begin{overpic}[scale=1]{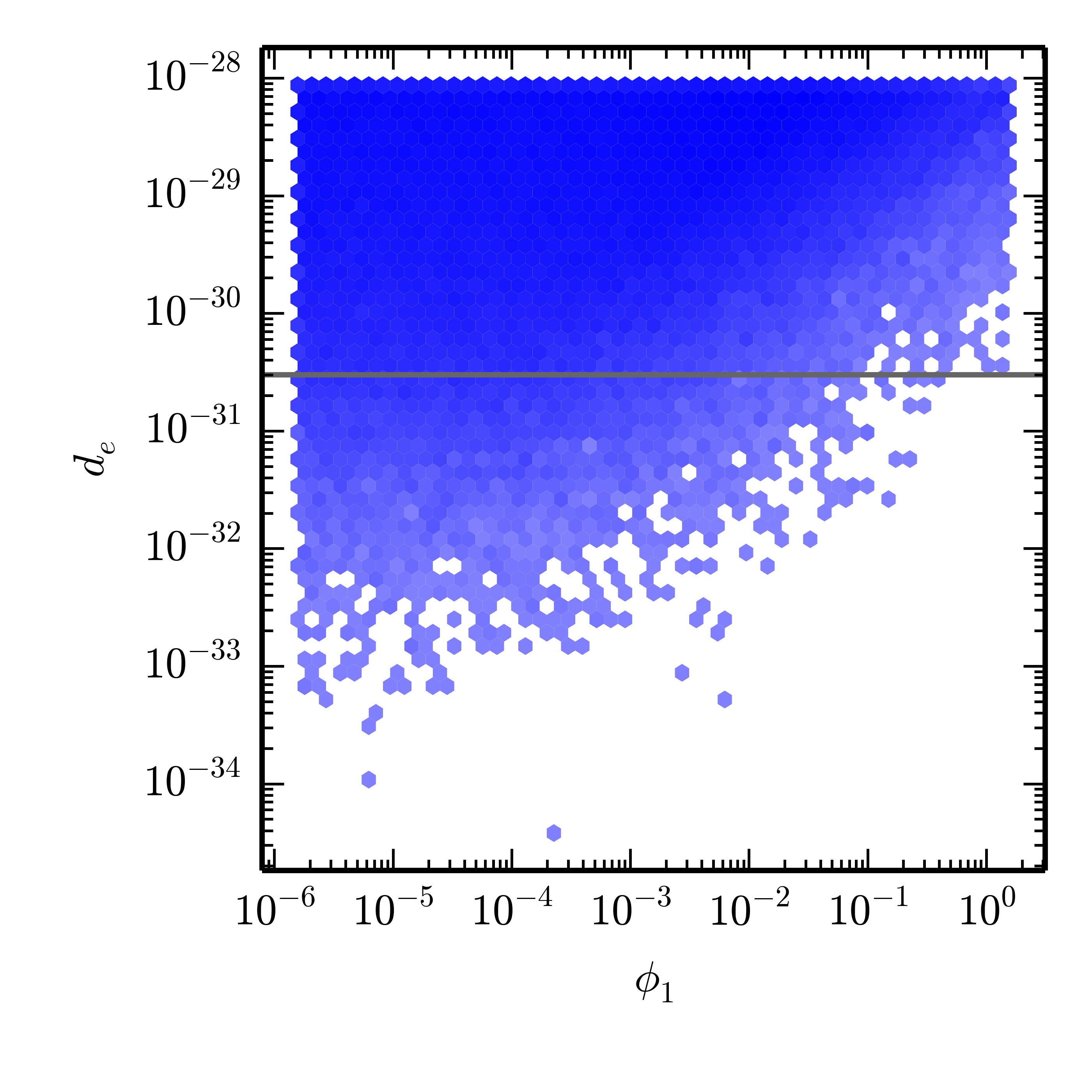} \put (80,30) {\Large } \end{overpic} &
\begin{overpic}[scale=1]{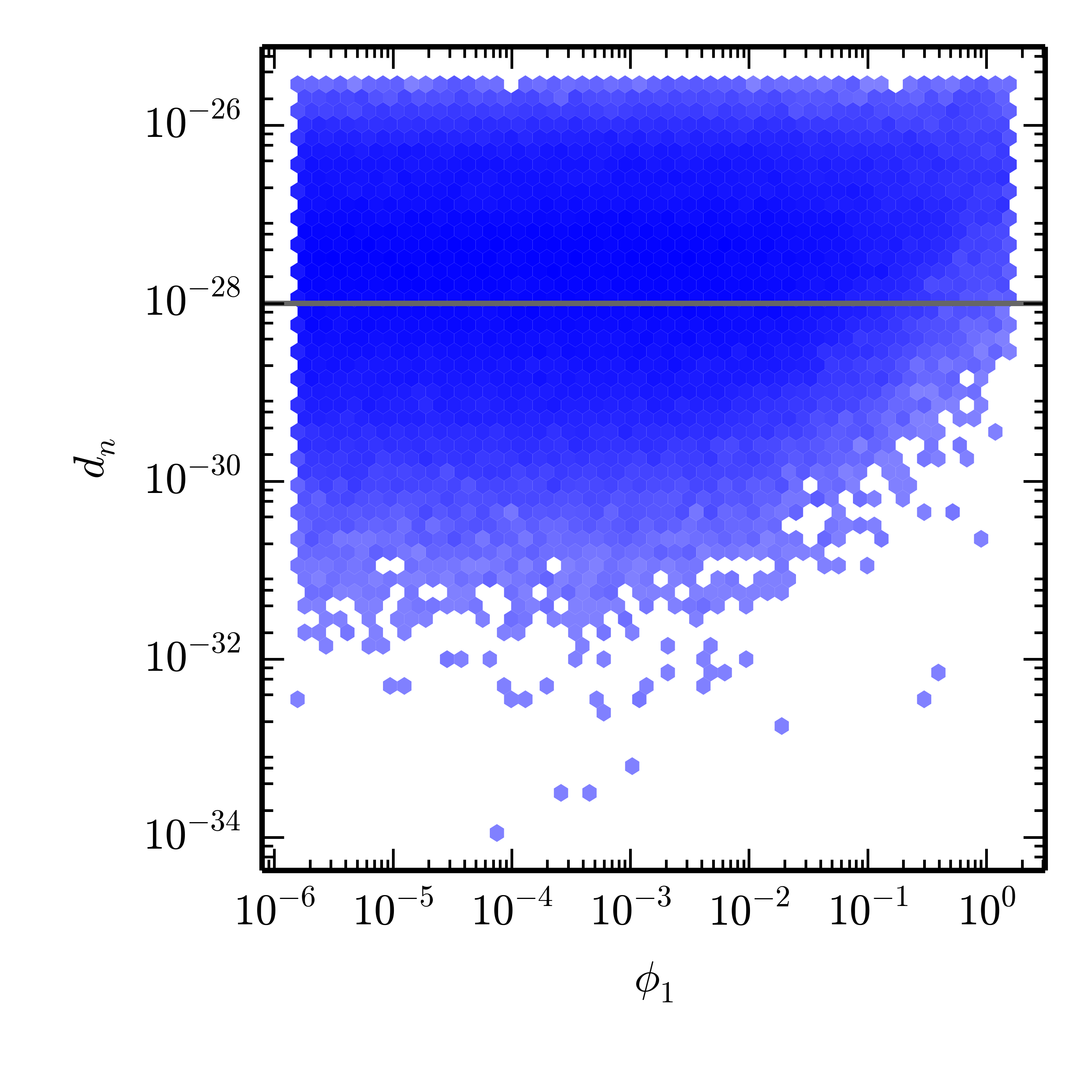} \put (80,30) {\Large } \end{overpic} \\
\begin{overpic}[scale=1]{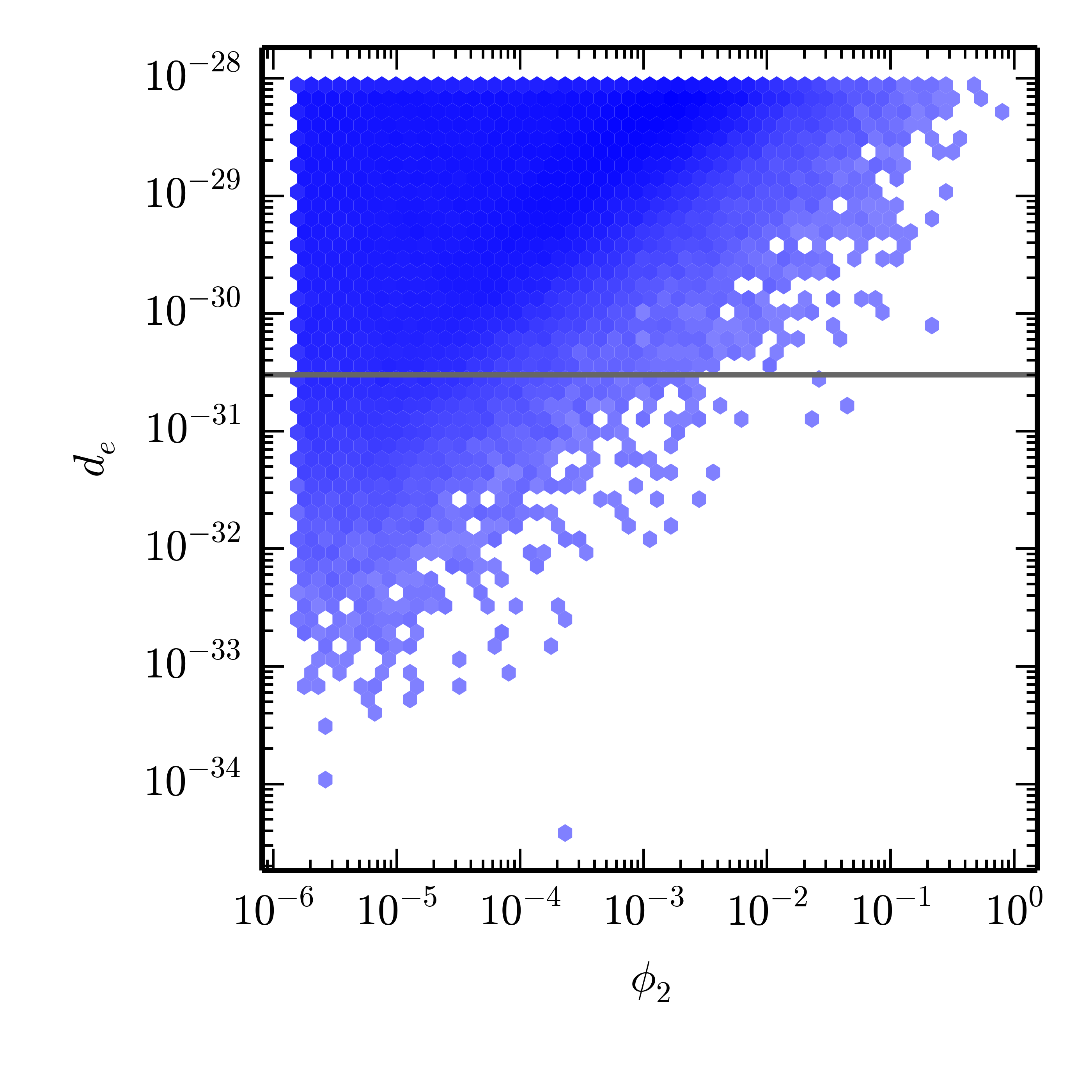} \put (80,30) {\Large } \end{overpic} &
\begin{overpic}[scale=1]{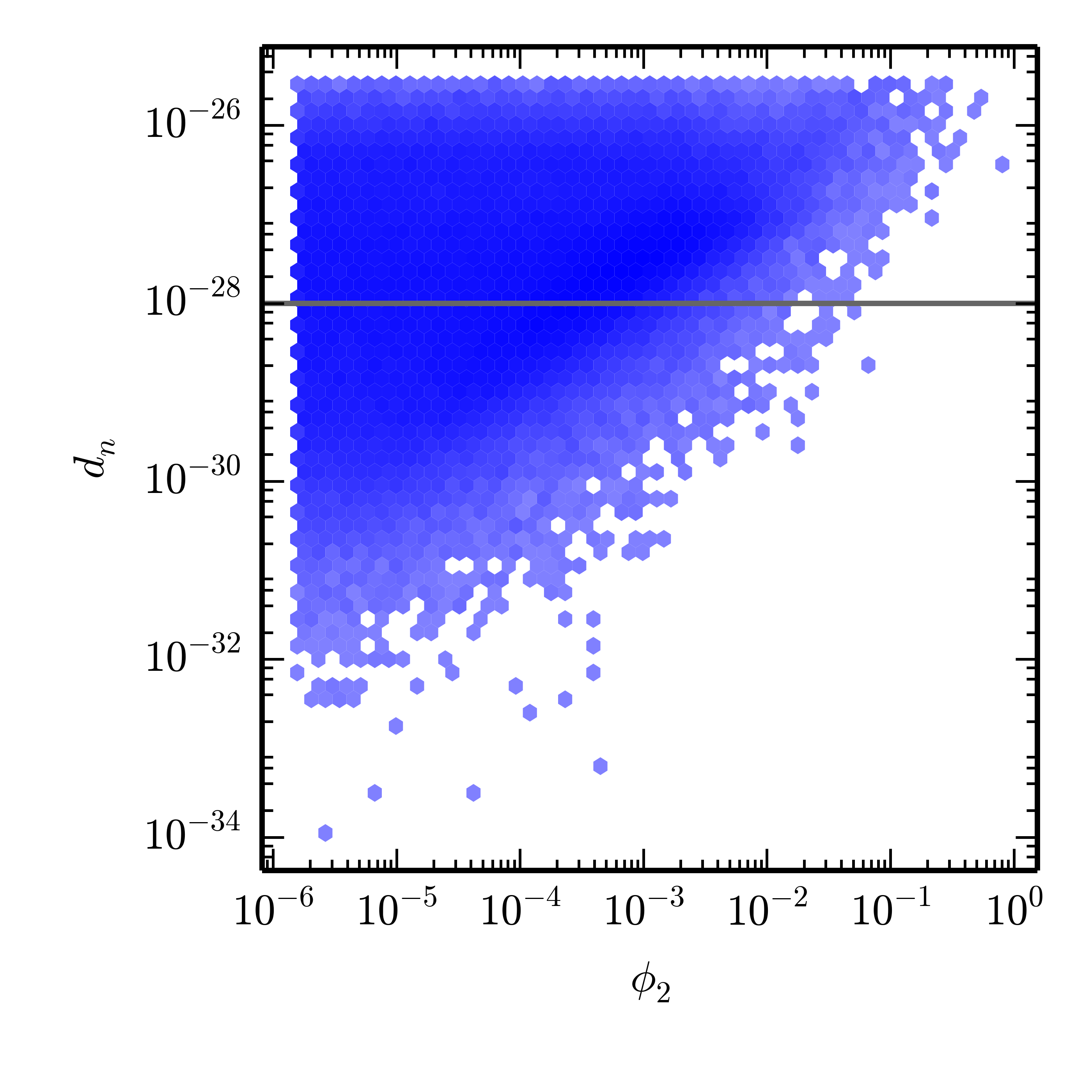} \put (80,30) {\Large } \end{overpic} \\
\begin{overpic}[scale=1]{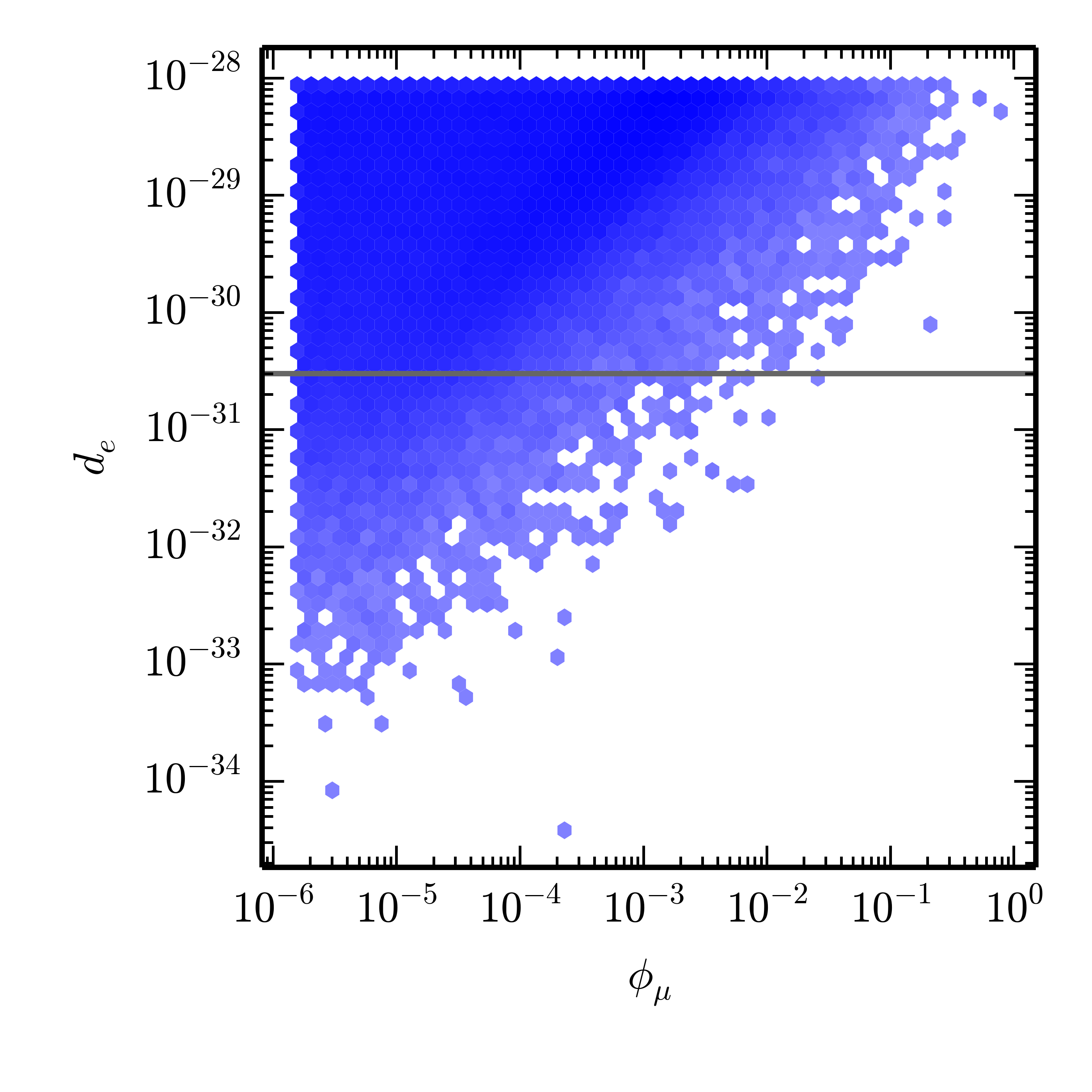} \put (80,30) {\Large } \end{overpic} &
\begin{overpic}[scale=1]{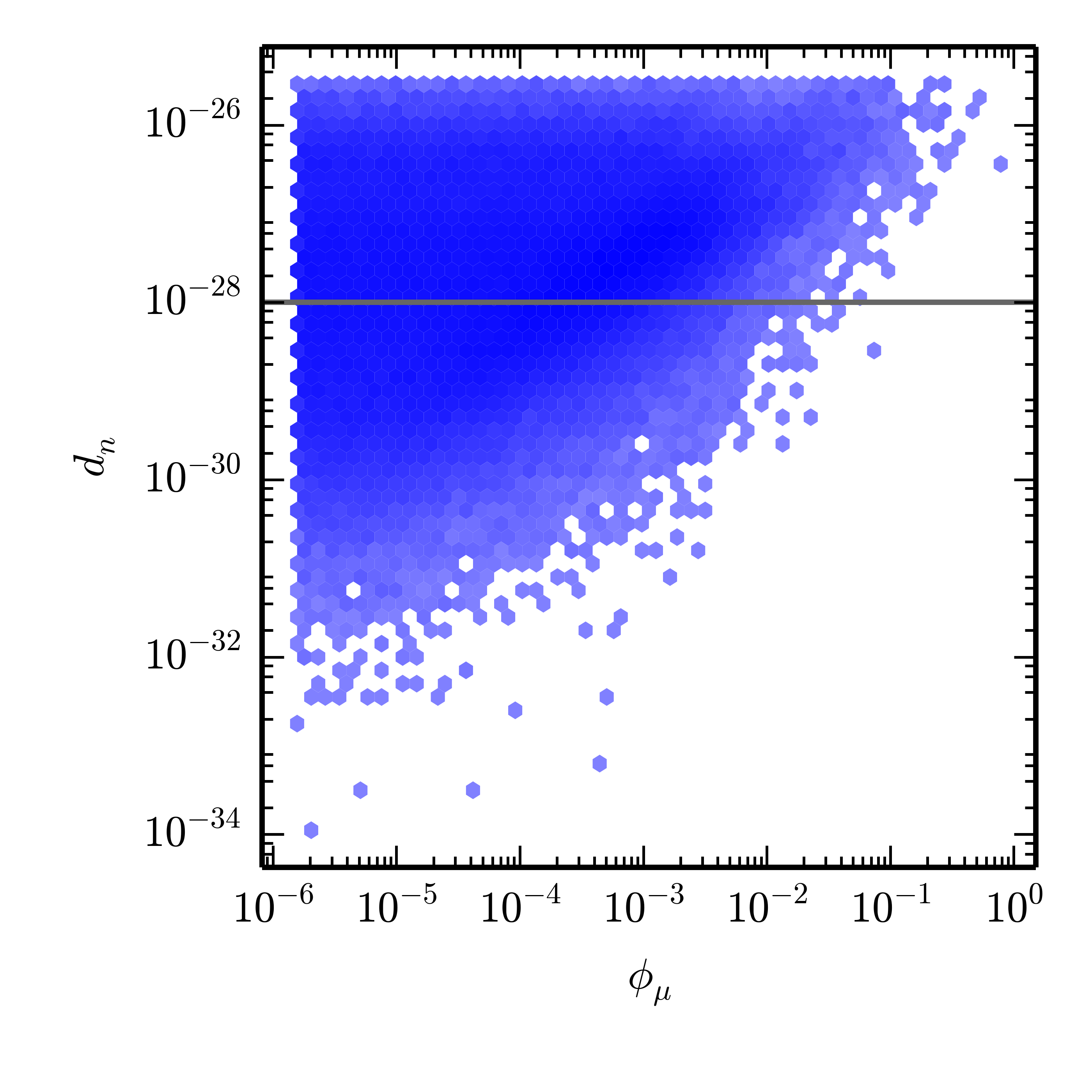} \put (80,30) {\Large } \end{overpic}
\end{tabular}
\caption[]{Model densities for the electron and neutron EDMs as a function of the phases
$\phi_1$, $\phi_2$ and $\phi_\mu$ in model set C.  The shading is as in Figure
\ref{fig:1}(f). The lines indicate the future expected EDM reaches.
\label{fig:3}}
\end{center}
\end{figure}

To further explore the constraints placed on the phases, Figure \ref{fig:3} contains two-dimensional histograms showing the neutrino and 
electrons EDMs paired with $\phi_1$, $\phi_2$, or $\phi_\mu$ for models that are allowed by current constraints (model set C).  
The anticipated future EDM search reach corresponds to the horizontal line.  The electron EDM 
provides a particularly strong constraint on large values of all three phases.  Most models 
with phases larger than $\sim 10^{-1}$ are already excluded, and the parameter space will be further explored by an 
order of magnitude in the values of the phases to roughly $10^{-2}$.  However, some room remains for rather large 
values of the phases due to possible cancellations of the contributions arising from $\phi_\mu$ and $\phi_2$ as described 
above.

\begin{figure}[!ht]
\begin{center}
\begin{tabular}{cc}
\begin{overpic}[scale=1]{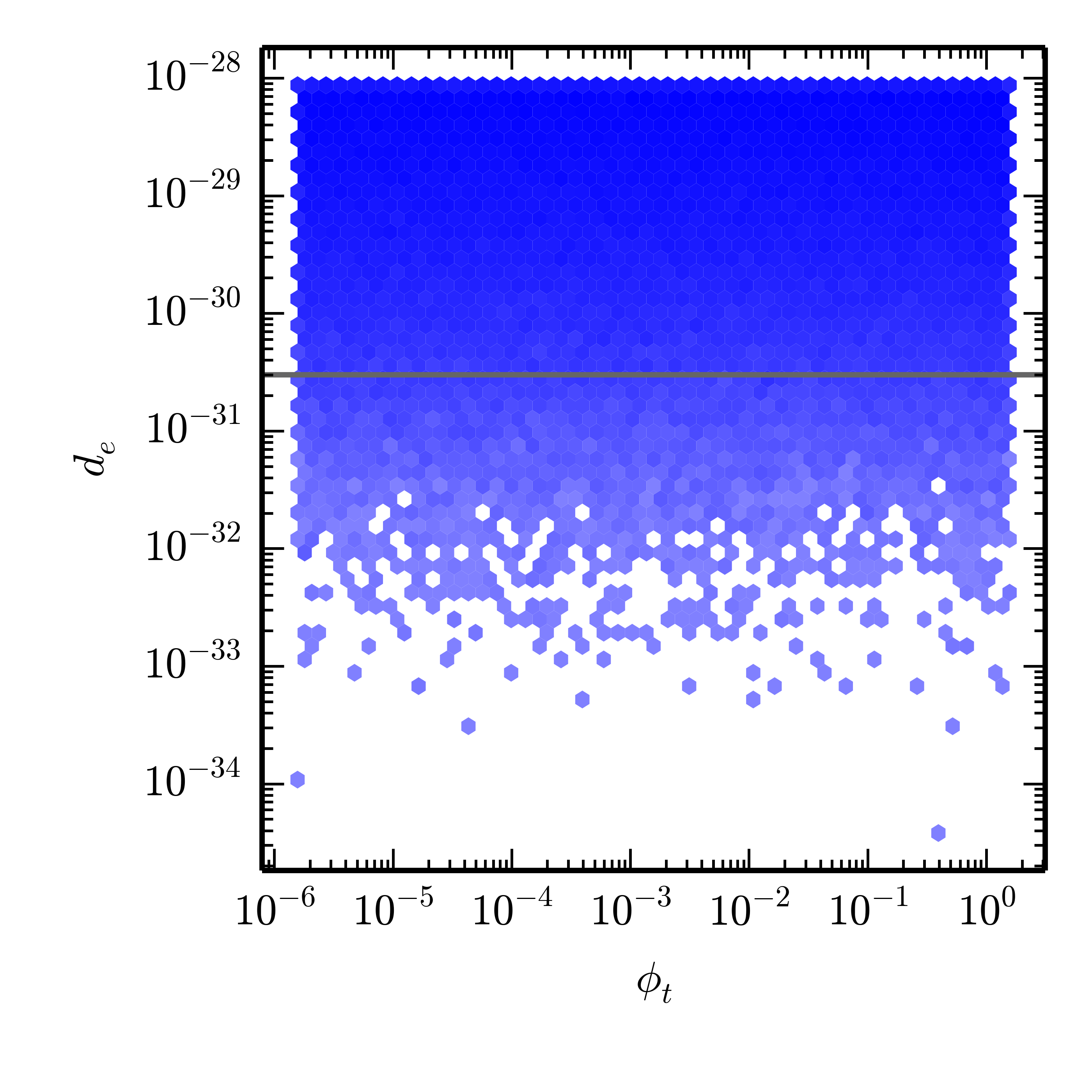} \put (40,25) {\Large} \end{overpic} &
\begin{overpic}[scale=1]{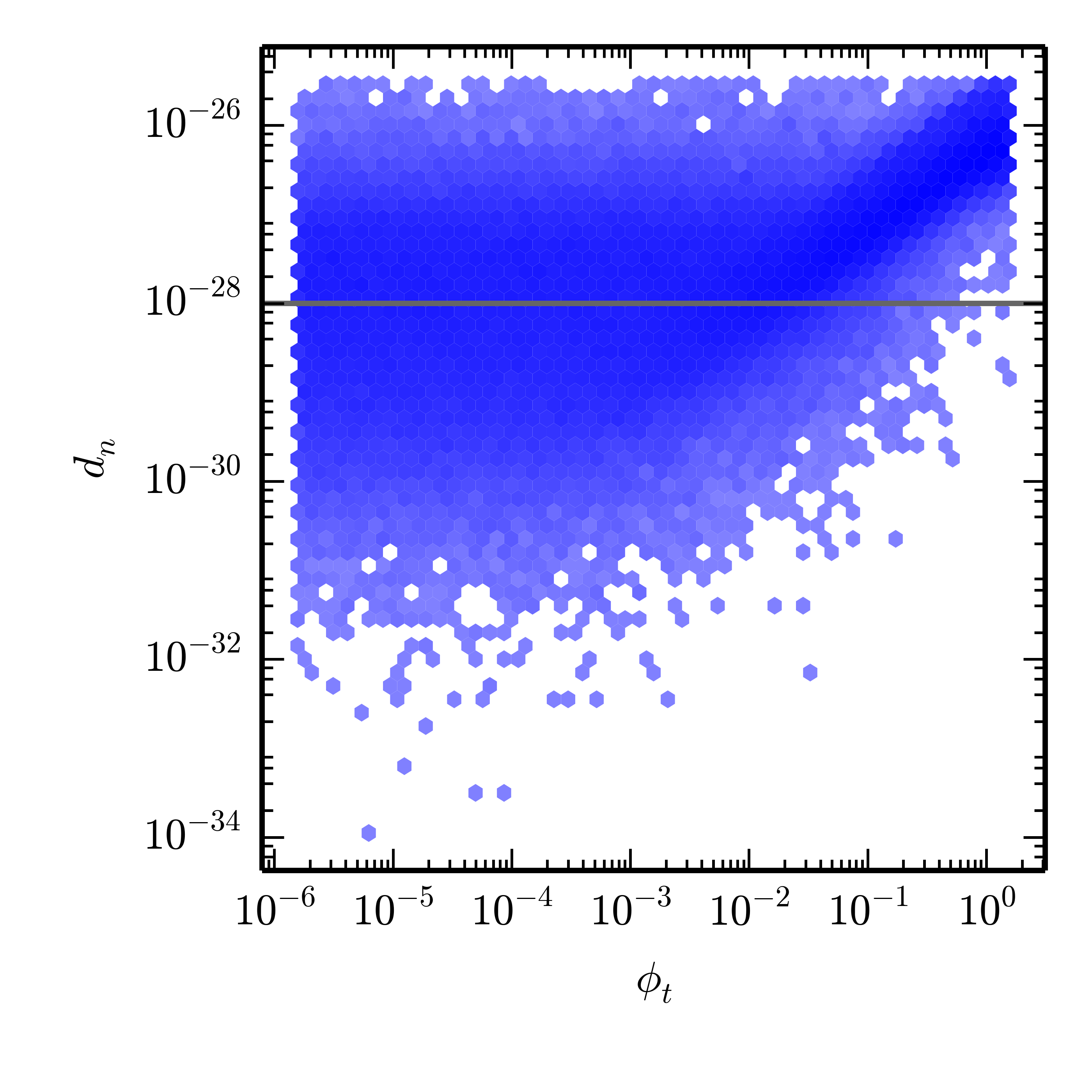} \put (80,30) {\Large} \end{overpic} \\
\begin{overpic}[scale=1]{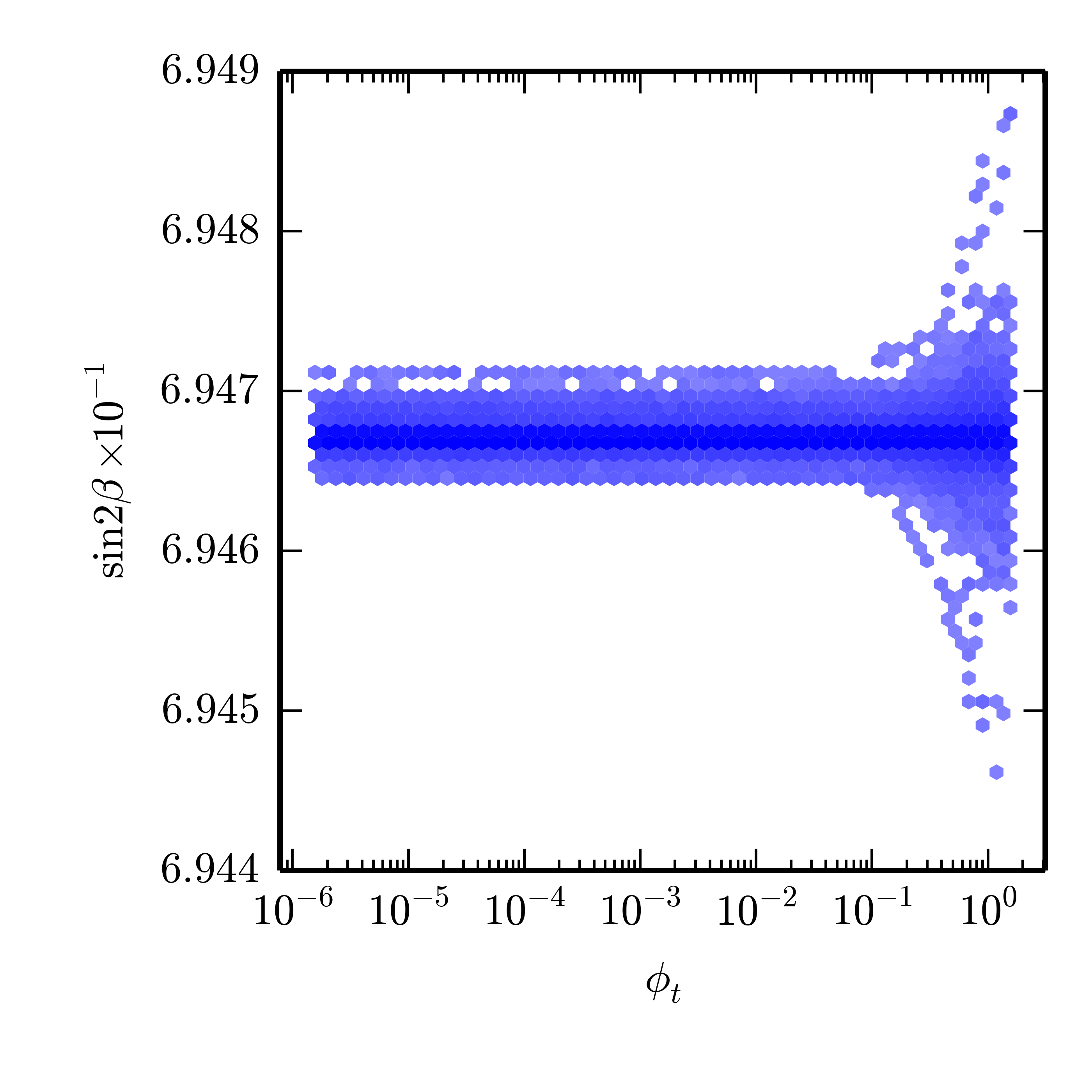} \put (30,30) {\Large} \end{overpic} &
\begin{overpic}[scale=1]{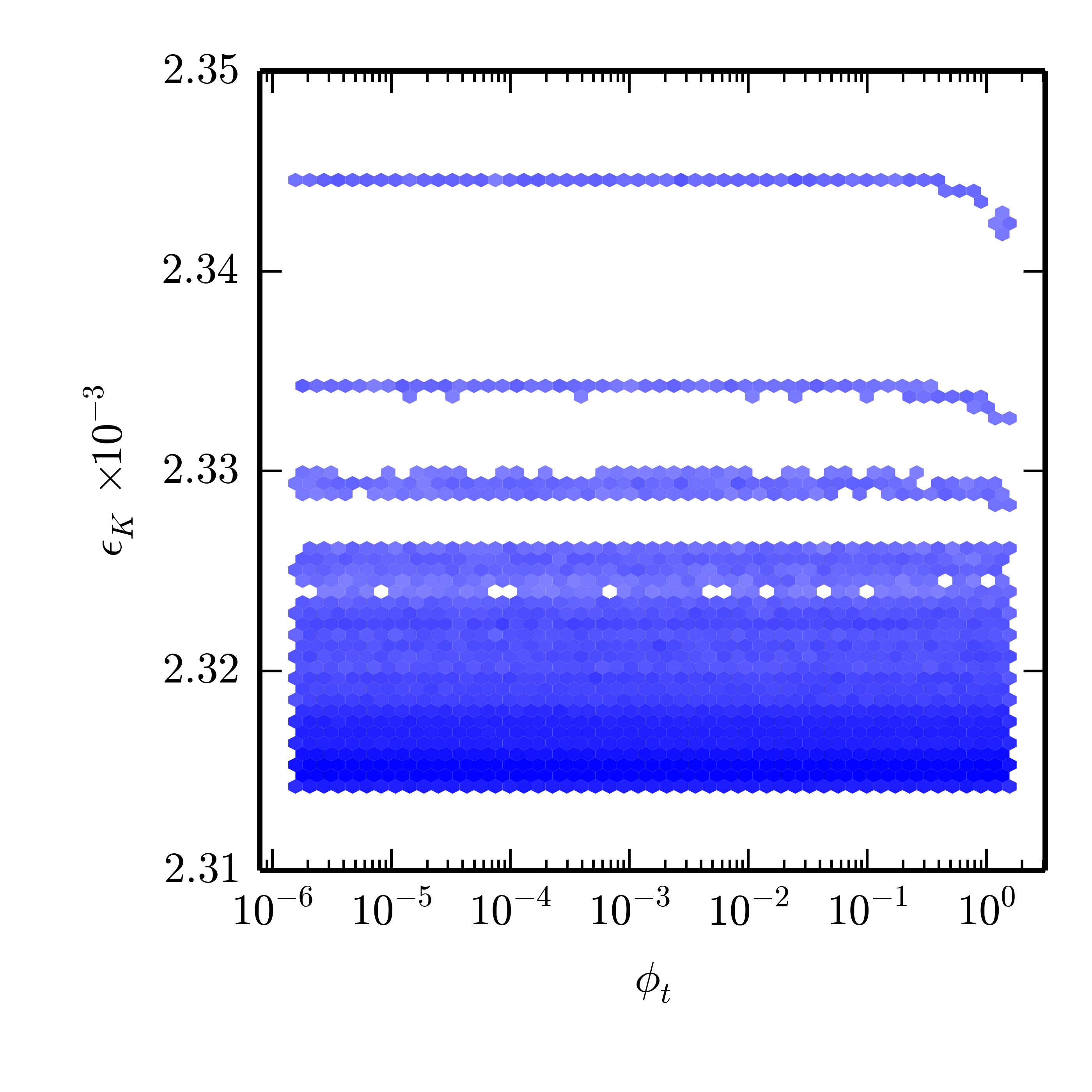} \put (30,75) {\Large} \end{overpic}
\end{tabular}
\caption[]{Model densities for the electron and neutron EDMs, $\sin2\beta$ and
  $\epsilon_K$ as a function of the phase $\phi_t$ in model set C.  The shading is as in
  Figure \ref{fig:1}(f).  The horizontal lines in the top panels indicate
  expected future EDM limits. The expected future constraints on $\sin2\beta$ and
  $\epsilon_K$ lie outside the ranges of the lower panels.
\label{fig:4}}
\end{center}
\end{figure}

Low-energy observables that are sensitive to the $A$-term phases $\phi_t$, $\phi_b$, and $\phi_\tau$
must involve flavor-violating couplings. The most sensitive such
observables  that are incorporated in \verb+SUSY_FLAVOR+ are $\sin
2\beta$ and $\epsilon_K$.  The dependence  
on $\phi_b$ and $\phi_\tau$ in these observables is extremely weak; $\phi_b$ 
contributions are typically suppressed by $m_b$ (except possibly at large $\tan\beta$), while 
$\phi_\tau$ would contribute only to $\tau$ flavor violating processes or poorly measured observables such as 
the $\tau$ EDM.  The impact of $\phi_t$ on the electron and neutron EDMs, as well as on $\sin 
2\beta$ and $\epsilon_K$, is shown in Figure \ref{fig:4}.  Unlike the structure illustrated in Figure 
\ref{fig:3}, we see that there is no correlation between most of these quantities and the phase $\phi_t$. We 
note that a mild constraint on models with large $\phi_t$ is possible from the neutron EDM, however
improving the precision on $\sin2\beta$ and $\epsilon_K$ does not bound these phases.

\begin{figure}[]
\begin{center}
\begin{tabular}{cc}
\begin{overpic}[scale=1]{hist_phi_2.png} \put (30,85) {\Large } \end{overpic} &
\begin{overpic}[scale=1]{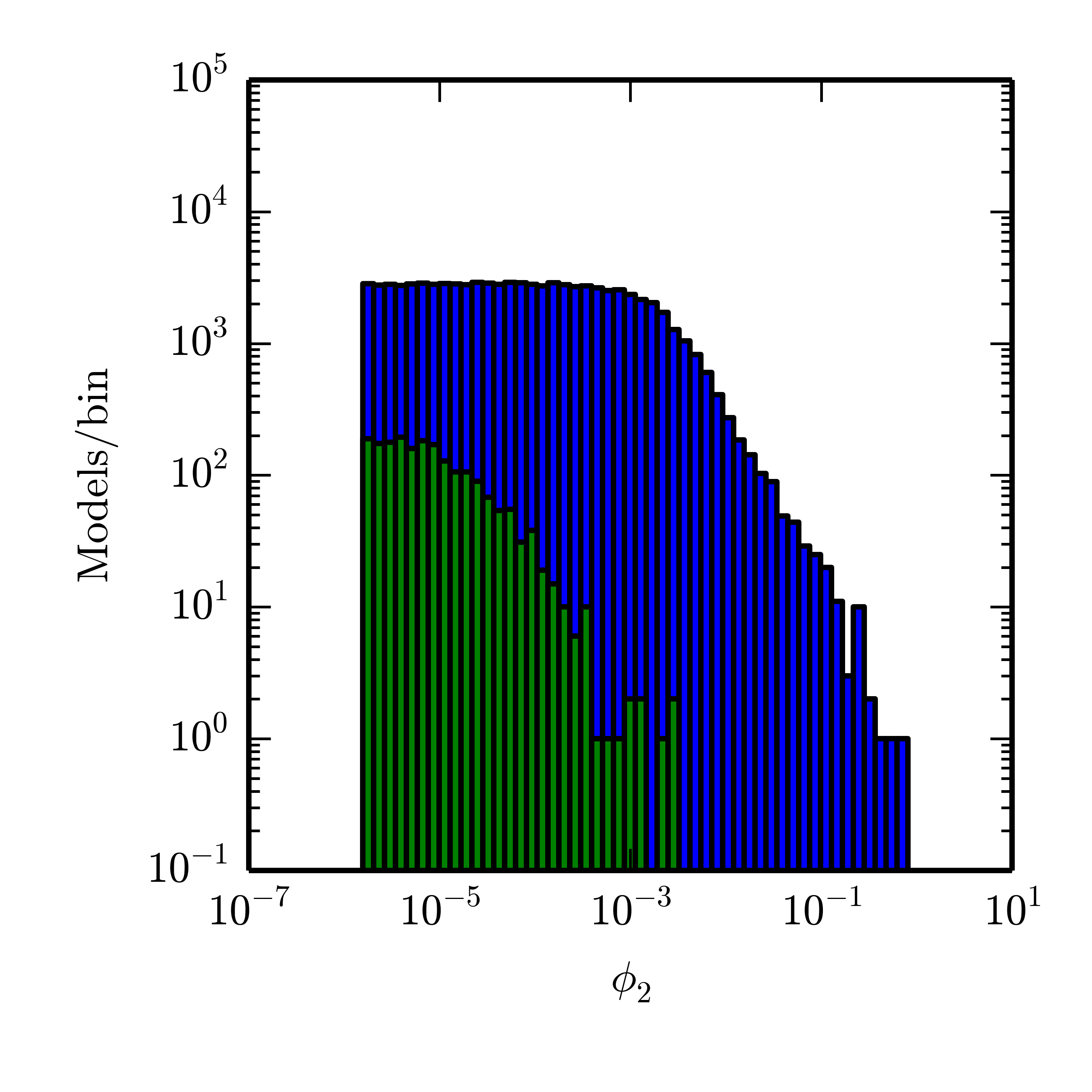} \put (30,85) {\Large } \end{overpic} \\
\begin{overpic}[scale=1]{hist_edm_e.png} \put (30,85) {\Large } \end{overpic} &
\begin{overpic}[scale=1]{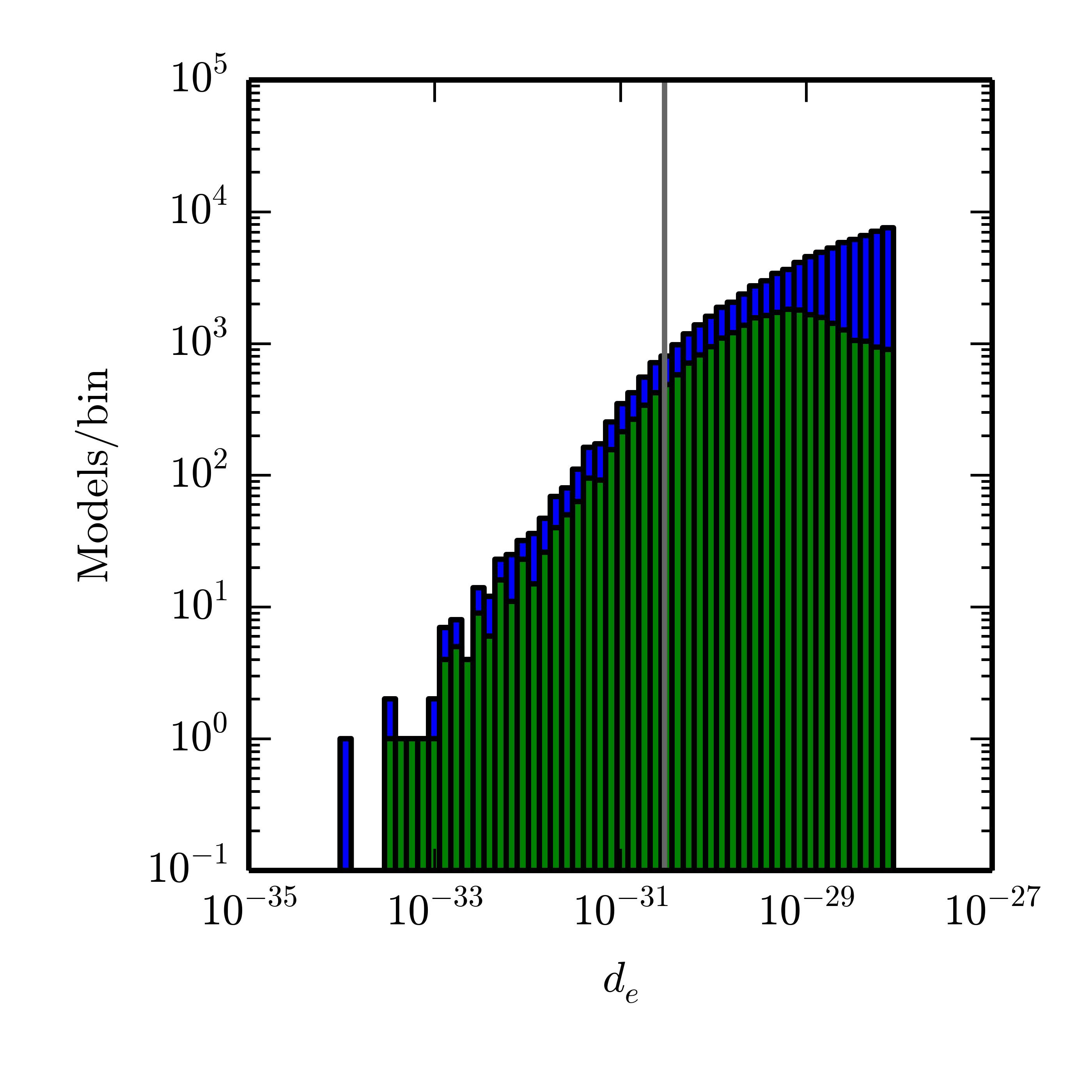} \put (30,85) {\Large } \end{overpic} \\
\begin{overpic}[scale=1]{hist_sin2b.png} \put (30,85) {\Large } \end{overpic} &
\begin{overpic}[scale=1]{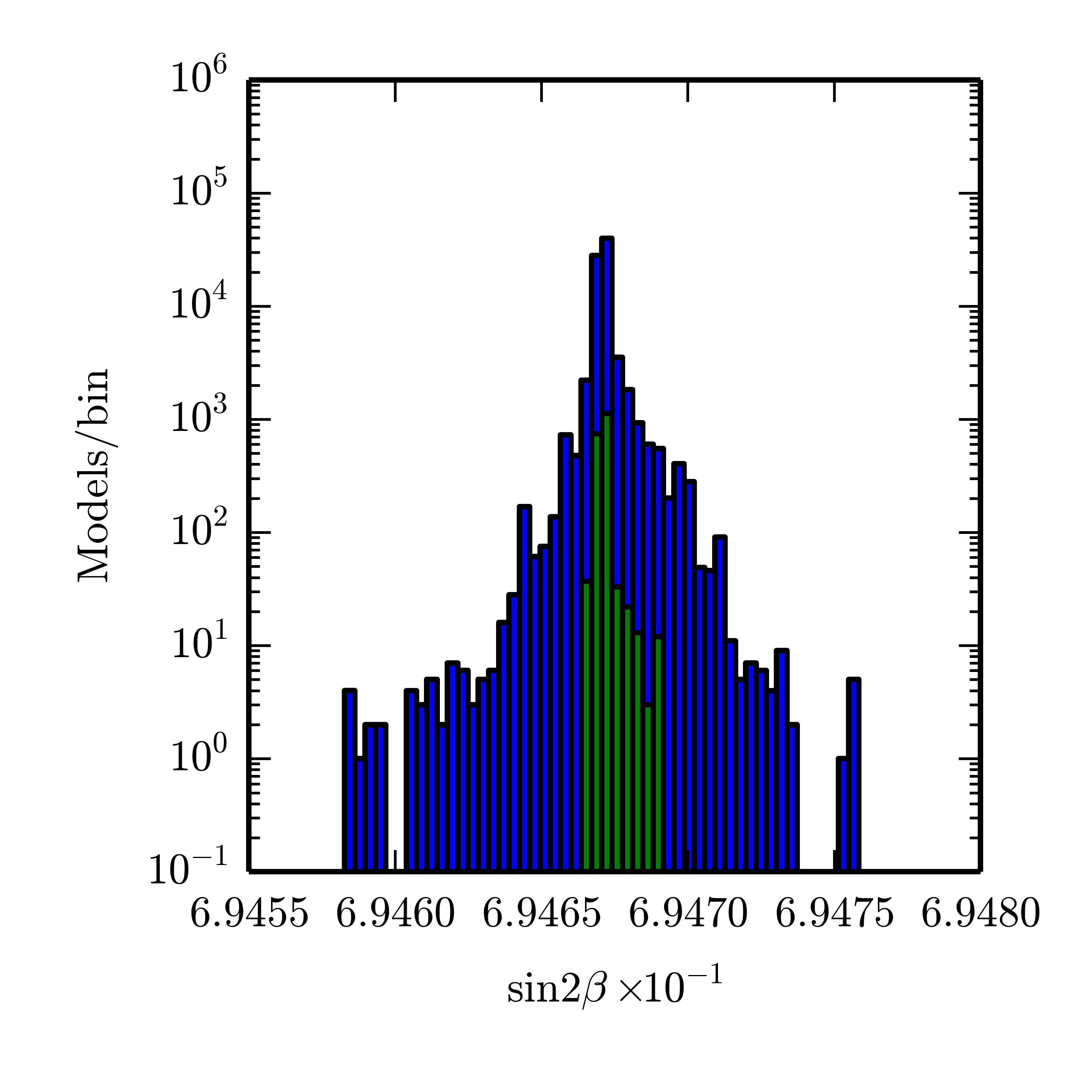} \put (30,85) {\Large } \end{overpic}\\[-0.5cm]
\end{tabular}
\caption[]{Complementarity between the low-energy experiments and the direct SUSY searches at 
the LHC. The left panels are identical to Figures \ref{fig:1}(a), \ref{fig:2}(a), and \ref{fig:2}(c), 
respectively. The right panels are the corresponding figures including only those models that are expected to 
evade the leading 14 TeV LHC direct SUSY searches with 3000 fb${}^{-1}$ of integrated luminosity.  Note the difference
in the scale of the vertical axis between the left and right panels.
\label{fig:5}}
\end{center}
\end{figure}

Next, we consider the correlation between low-energy observables and direct LHC searches for Supersymmetry in order to 
determine the degree of complementarity between the two approaches for probing Supersymmetry.  
To do so, we consider models that are expected \cite{Cahill-Rowley:2014twa}  
to evade Jets+MET and stop-squark searches at the LHC with
3000 fb${}^{-1}$ of integrated luminosity (extracted from model sets C and D). Figure \ref{fig:5} shows a comparison 
of the impact of low-energy experiments to that of direct LHC searches in exploring the
pMSSM models with CP phases. In all cases, we see that the shape of the distribution remains essentially unchanged,
and only the number of viable models is affected by the LHC searches.  Hence, 
the ability of the current and future measurements of low-energy observables to constrain models 
is observed to be independent of the the discovery reach of the LHC.  These results indicate a high degree 
of complementarity between the low-energy CP violation experiments and the direct SUSY searches 
at the LHC as probes of the MSSM.
\clearpage
%
%
\clearpage
%

\begin{figure}[]
\begin{center}
\begin{tabular}{cc}
\begin{overpic}[scale=1]{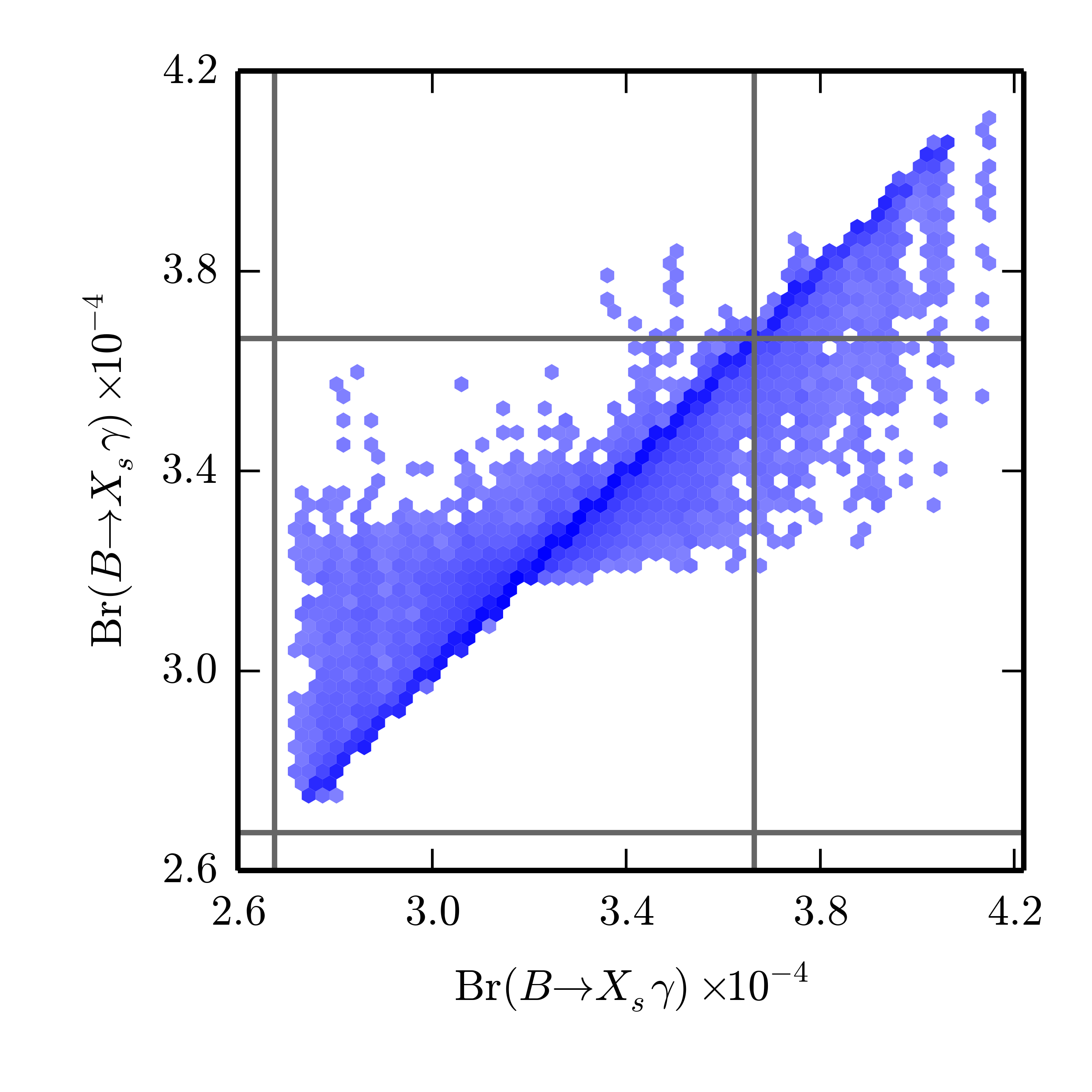} \put (80,30) {\Large } \end{overpic} &
\begin{overpic}[scale=1]{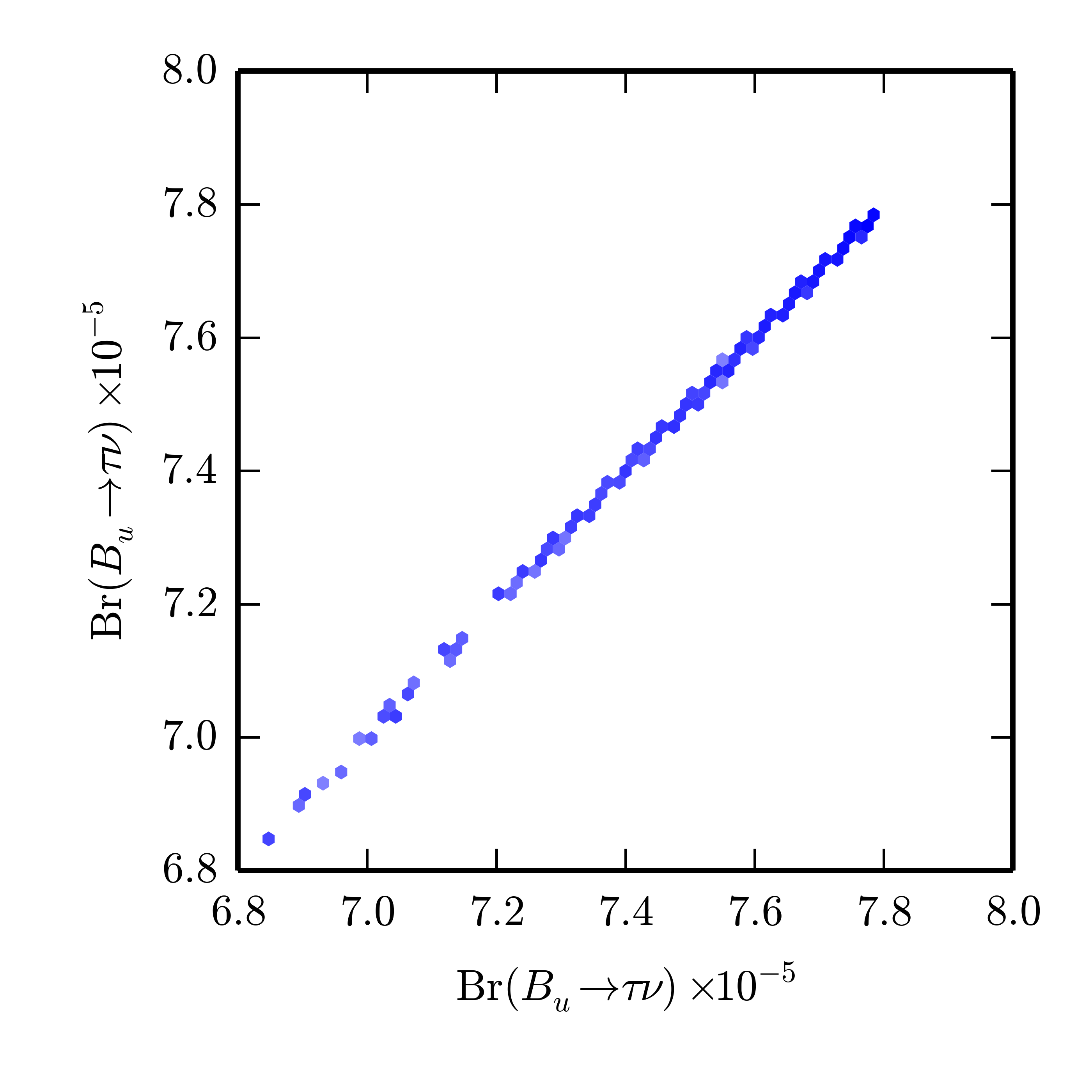} \put (80,30) {\Large } \end{overpic} \\[-0.5cm]
\begin{overpic}[scale=1]{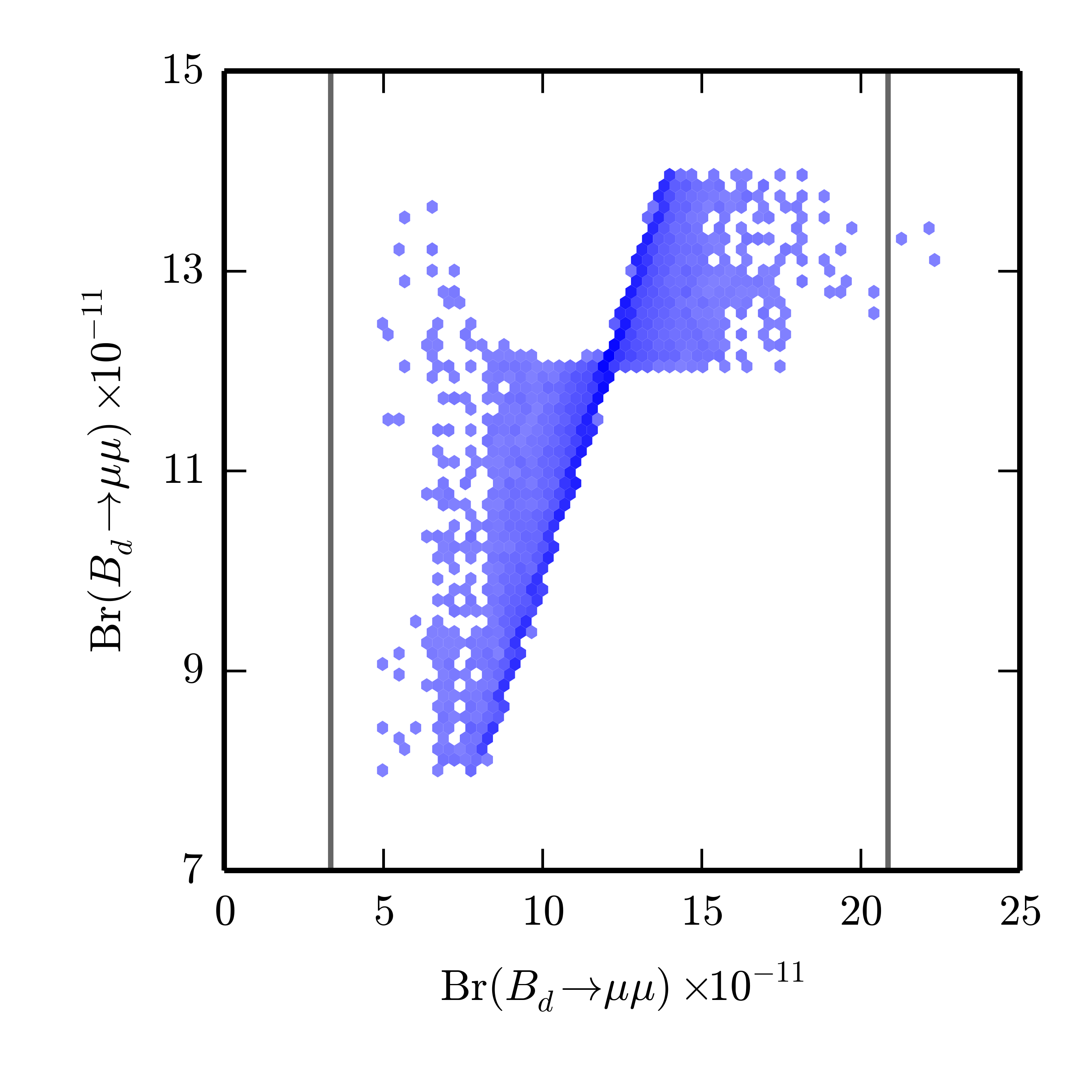} \put (80,33) {\Large } \end{overpic} &
\begin{overpic}[scale=1]{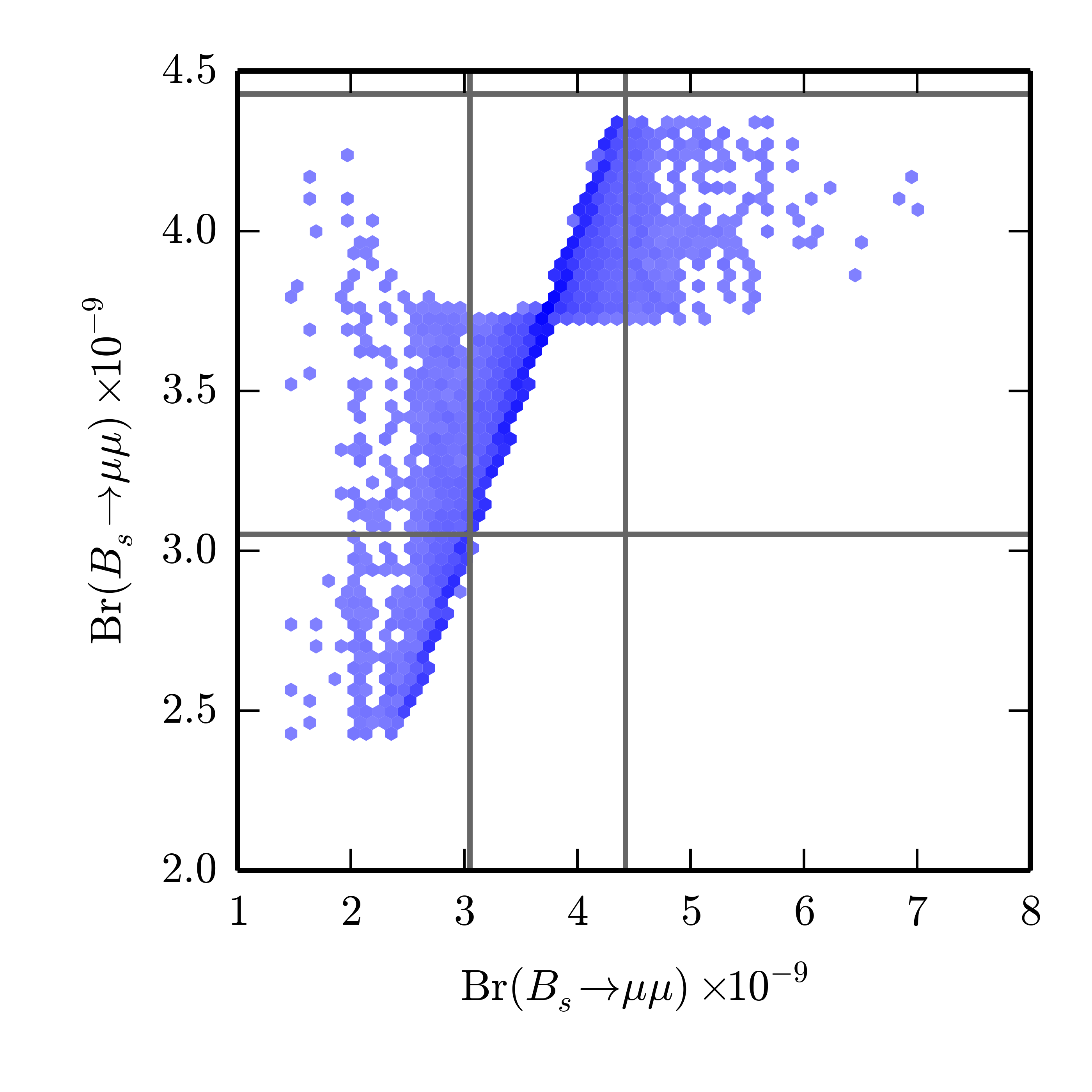} \put (80,30) {\Large } \end{overpic} \\[-0.5cm]
\begin{overpic}[scale=1]{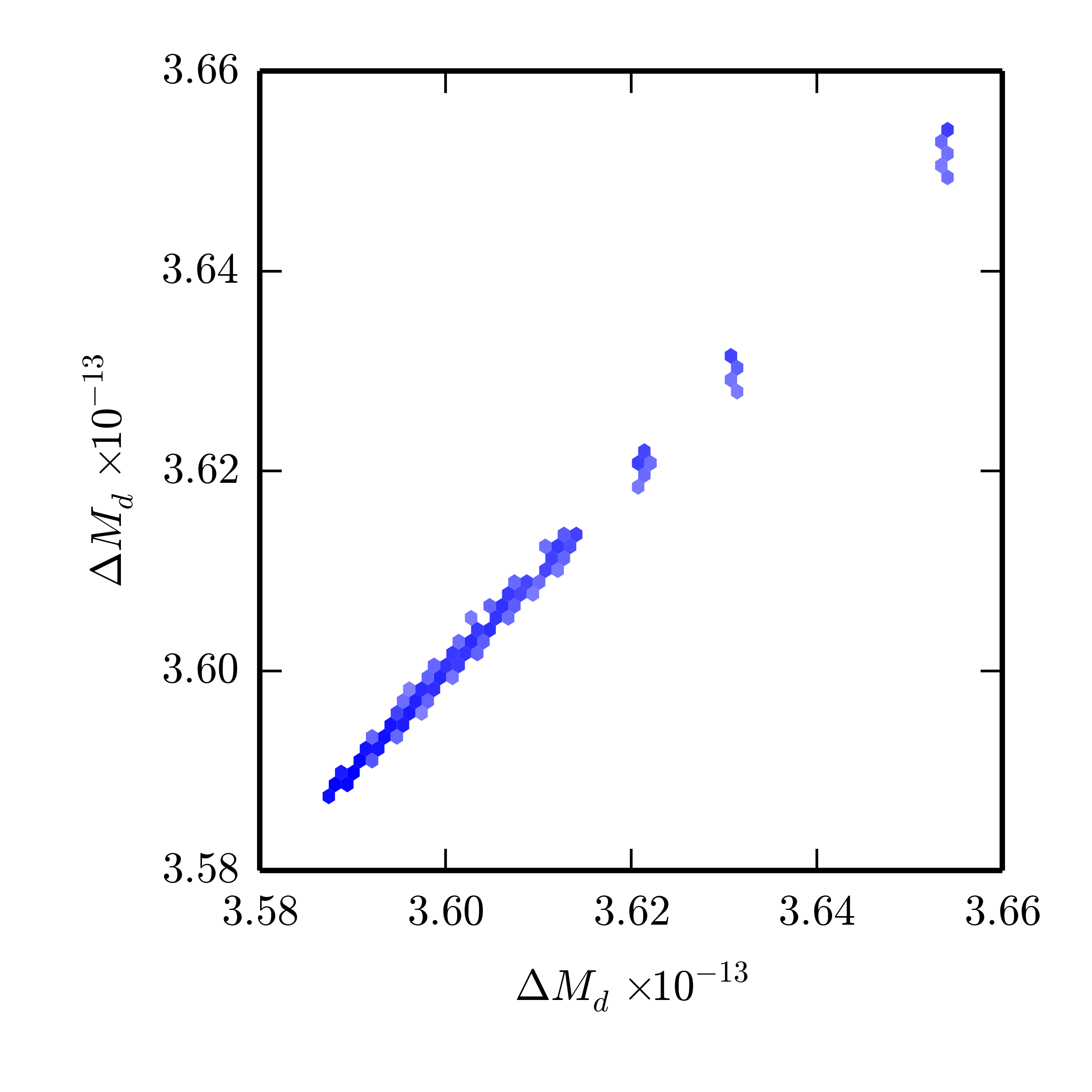} \put (80,30) {\Large } \end{overpic} &
\begin{overpic}[scale=1]{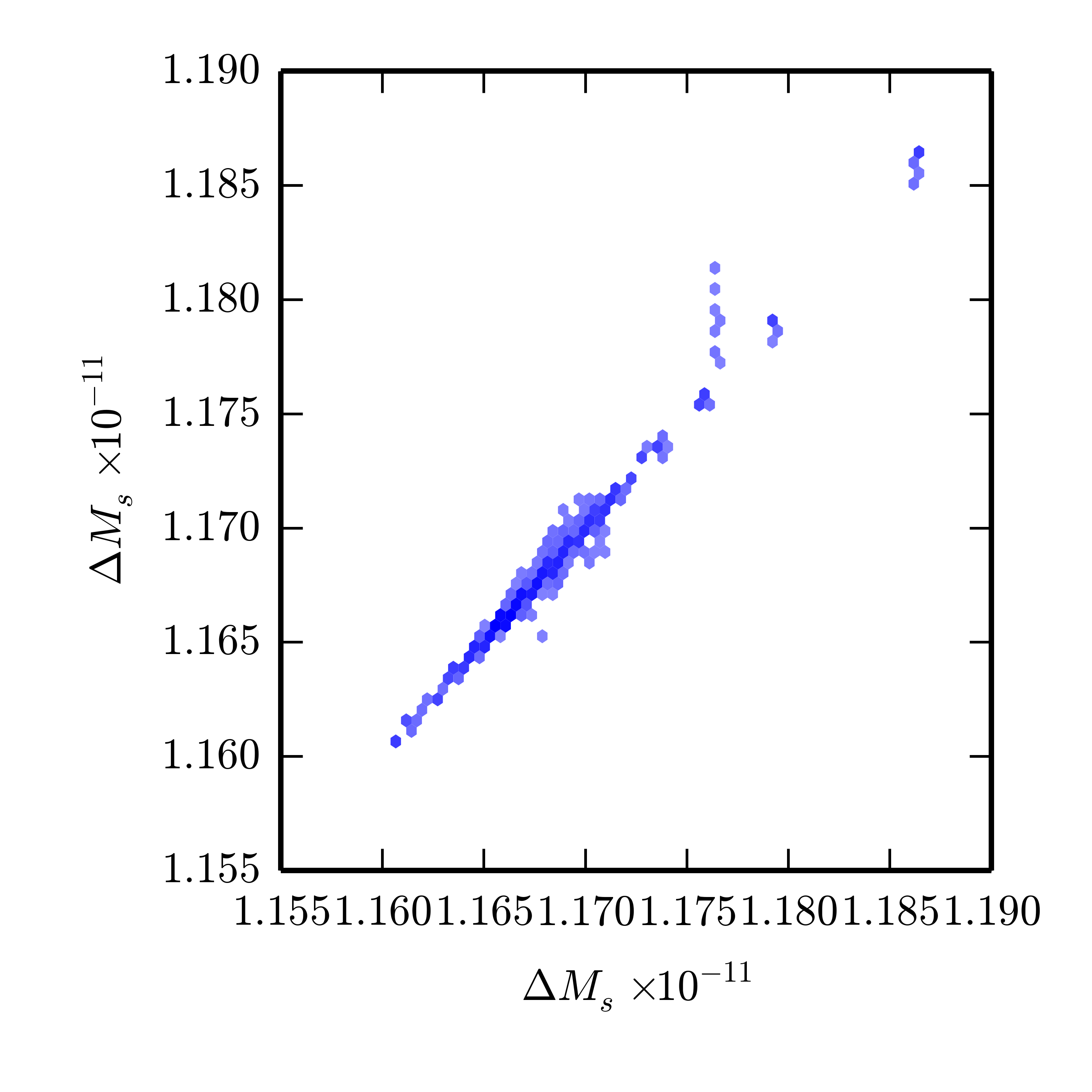} \put (80,30) {\Large } \end{overpic}\\[-0.5cm]
\end{tabular}
\caption[]{Comparison of several flavor-changing, CP-conserving $B$ physics observables in the original pMSSM
models without CP violation (model set A on the horizontal axes) and the corresponding 
models with randomly selected phases (model set C on the vertical axes). The lines indicate
future expected measurements on the various observables. The shading is as in
Figure \ref{fig:1}(f).
\label{fig:6}}
\end{center}
\end{figure}

\begin{figure}[!hb]
\begin{center}
\begin{tabular}{cc}
\begin{overpic}[scale=1]{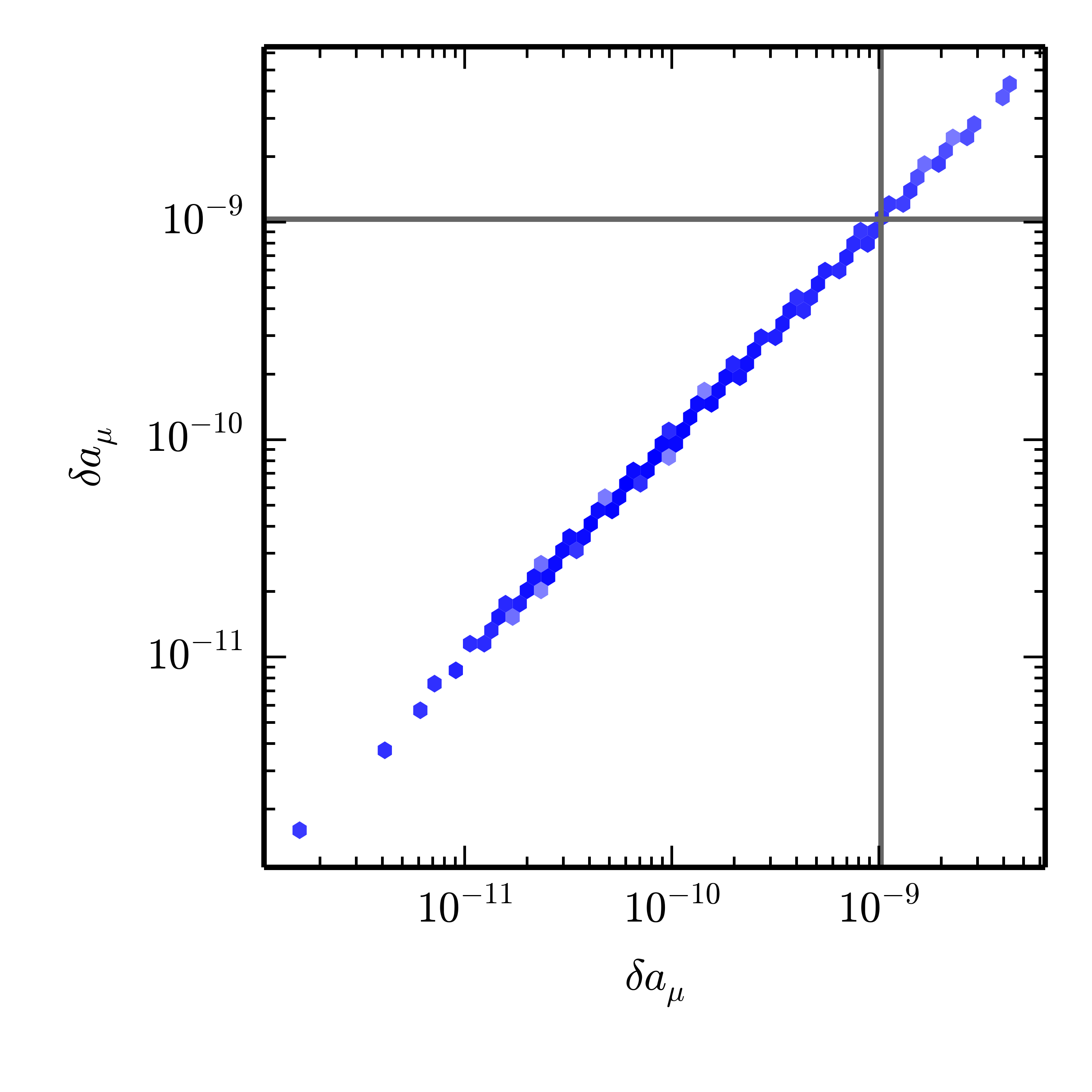} \put (80,30) {\Large} \end{overpic} &
\begin{overpic}[scale=1]{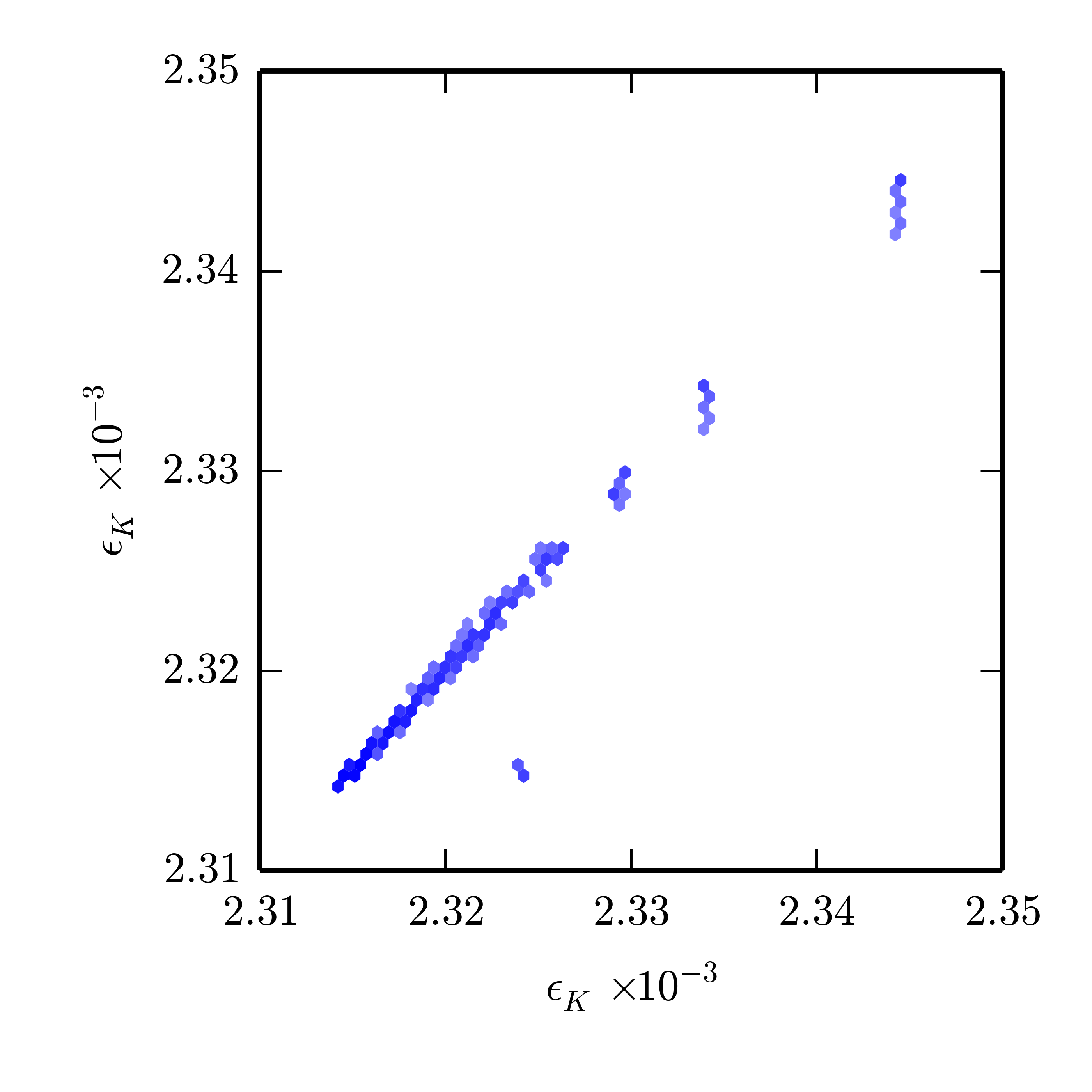} \put (80,30) {\Large} \end{overpic} \\[2mm]
\begin{overpic}[scale=1]{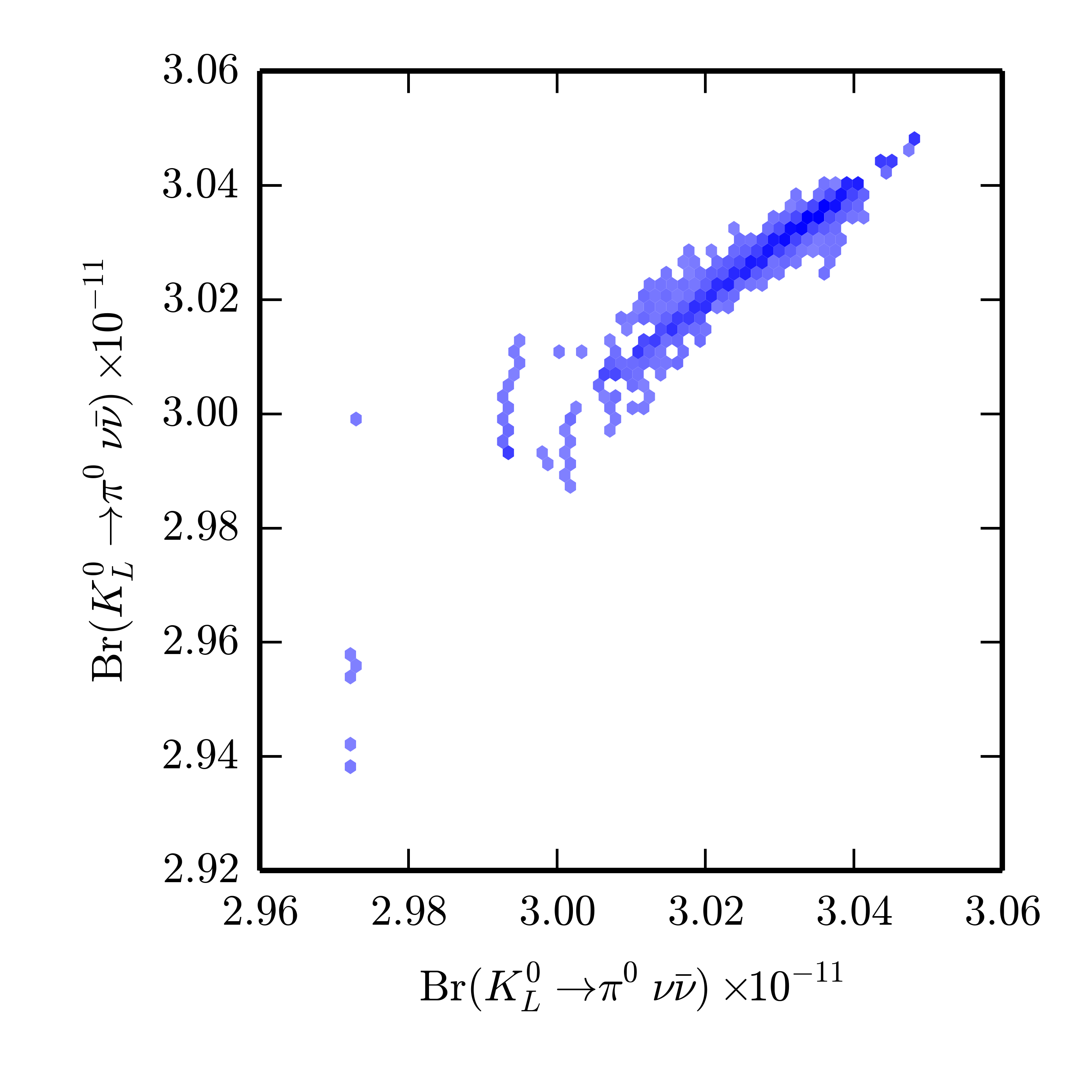} \put (80,30) {\Large} \end{overpic} &
\begin{overpic}[scale=1]{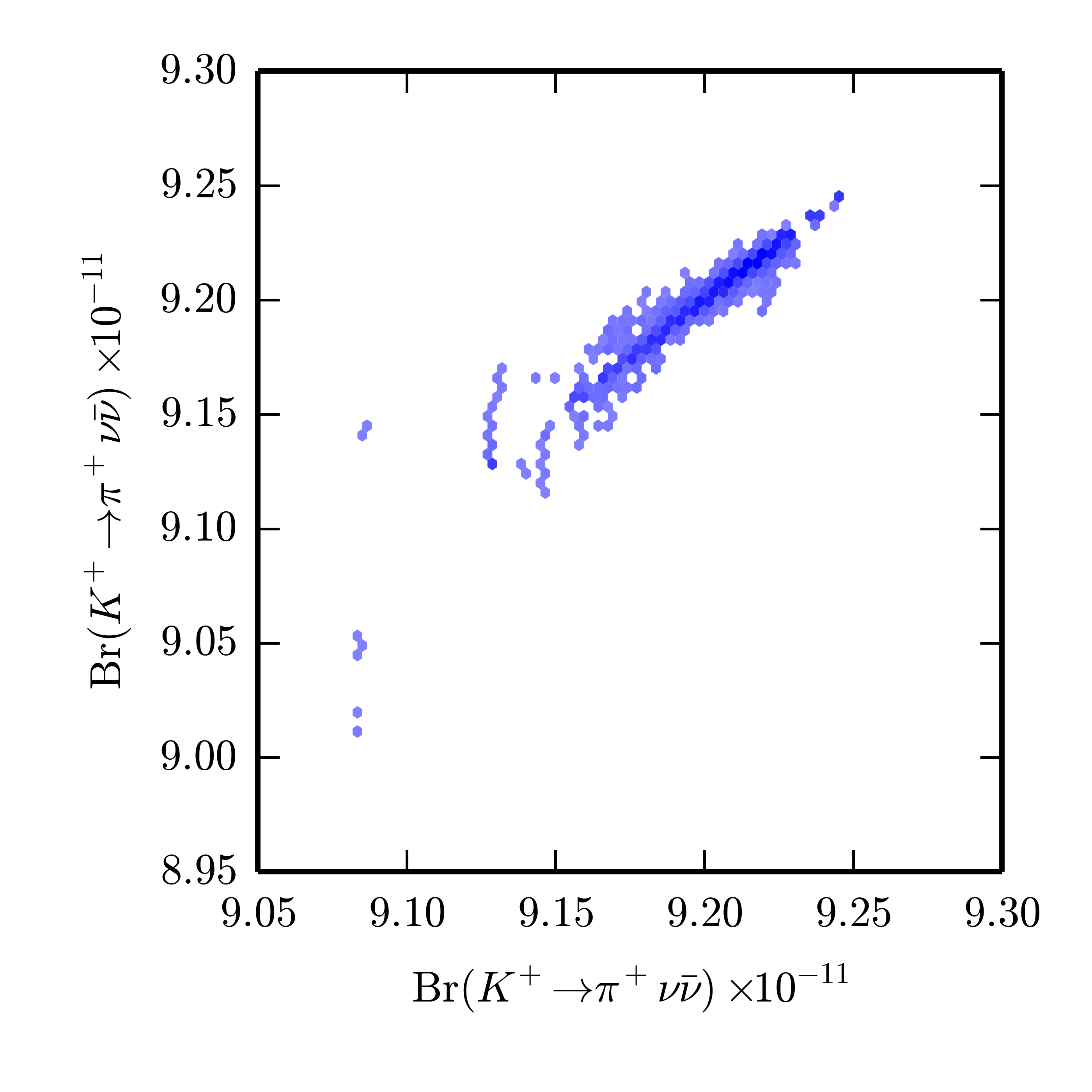} \put (80,30) {\Large} \end{overpic}\\[2mm]
\end{tabular}
\caption[]{Same as Figure \ref{fig:6}, but for the anomalous magnetic moment of the
muon, as well as rare Kaon processes.
\label{fig:7}}
\end{center}
\end{figure}

Although the main thrust of this work is to study CP-violating observables, flavor-changing CP-conserving observables, 
as well as the anomalous magnetic moment of the muon, are also sensitive to the presence of  
CP-violating phases.  We therefore next examine 
the most sensitive flavor-violating, CP-conserving observables to study 
their sensitivity to the pMSSM phases.  In Figures \ref{fig:6} and \ref{fig:7}, we compare the predicted rates of various 
processes in the original CP-conserving pMSSM (model set A on the x-axis) with their 
CP-violating counterparts (model set C on the y-axis).  While we see 
a strong correlation between the two model sets in the CP-conserving processes, it is clear that the presence of
phases lead to some interesting effects.
\clearpage
%
%
%
\clearpage
In particular, ${\rm Br}(B_{s,d} \to \mu^+ \mu^-)$ and ${\rm Br}(B \to X_s \gamma)$ all experience 
non-trivial contributions from supersymmetric CP phases and 
future experiments should have a significant sensitivity to these contributions. It is 
particularly noteworthy that while some of the CP-conserving pMSSM models are expected to 
be probed by future measurements of ${\rm Br}(B \to X_s \gamma)$ and ${\rm Br}(B_s \to \mu^+ \mu^-)$, 
introducing CP-violating phases has the potential to bring predictions for these processes 
\emph{back into agreement} with the SM rates. 

In particular, the ratio of  ${\rm Br}(B_s \to \mu^+ \mu^-)$ in the MSSM to that in the SM can be written 
to a very good approximation in the following way, 
\begin{eqnarray}
\dfrac{{\rm Br}(B_s \to \mu^+ \mu^-)|_{\rm MSSM}}{{\rm Br}(B_s \to \mu^+ \mu^-)|_{\rm SM}}   =     |X|^2   + | 1 + X |^2 \, ,
\label{bsmumu-analytic}
\end{eqnarray}
where $X$ is the new contribution from the supersymmetric partners. Writing $X$ as $X= \pm x  \, e^ {i \, \theta}$,  Eq.~\eqref{bsmumu-analytic} 
reduces to (here $x$ is taken to be positive by convention and, barring large cancellation, must satisfy $x << 1$ to be consistent 
with experiment)
\begin{eqnarray}
\dfrac{{\rm Br}(B_s \to \mu^+ \mu^-)|_{\rm MSSM}}{{\rm Br}(B_s \to \mu^+ \mu^-)|_{\rm SM}}   &=&     x^2   + | 1 \pm x  \, e^ {i \, \theta}|^2 \\
&=& 1  \pm  2 x \cos\theta + 2 x^2 \, .
\end{eqnarray}
Clearly, for a given value of $x$, the above ratio maximally deviates from unity at $\theta =0$ and any finite non-zero value of the 
phase $\theta$ will always push it closer to the SM. Note that the above argument is strictly correct only in the case when the NP amplitude 
has single contribution. However, even in the presence of multiple contributions, if one of them is the dominant term (which is indeed the case 
here because the higgsino contribution dominates in most of the parameter space where the constraint from ${\rm Br}(B_s \to \mu^+ \mu^-)$ 
is important) or there are no large cancellations among the various terms, the above argument still holds.

We also see 
some cases where the addition of CP phases creates a signal in one of these 
flavor-changing processes that would be detectable by future experiments; this effect, however, is found to be far less common.
Note that the rare Kaon processes generally tend to not exhibit these effects and thus are much less sensitive to the
presence of CP phases.

\begin{figure}[!ht]
\begin{center}
\begin{tabular}{cc}
\includegraphics[scale=1]{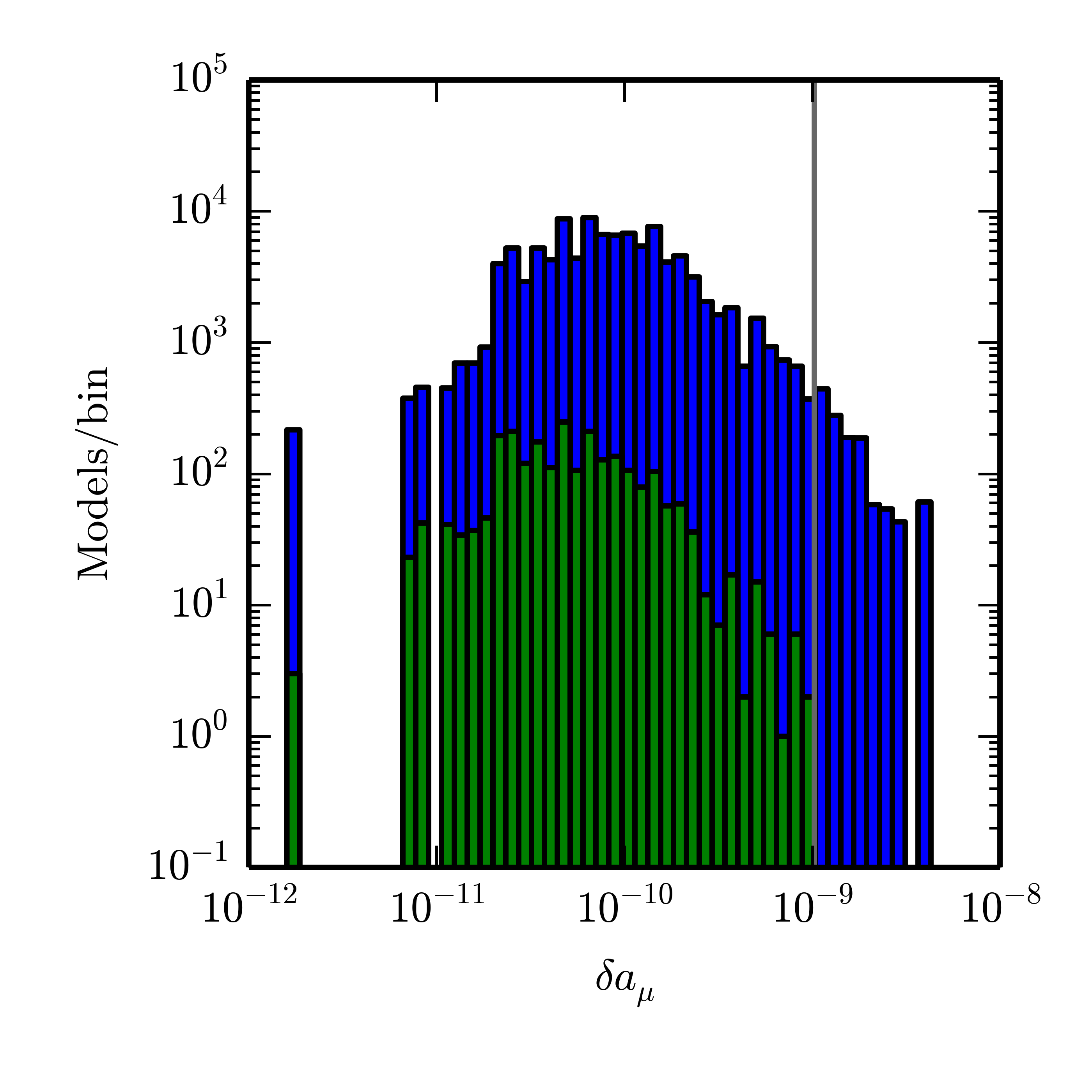} &
\includegraphics[scale=1]{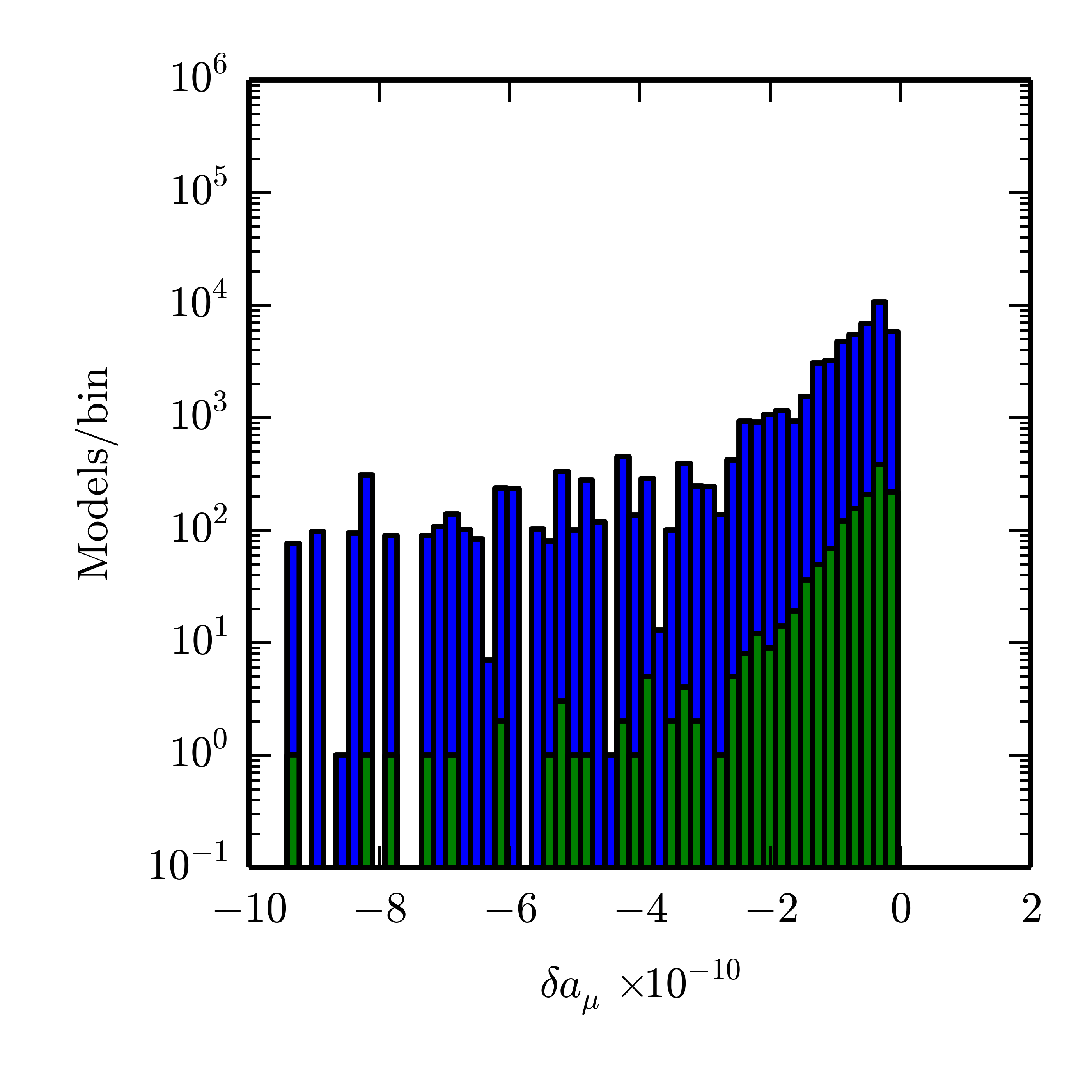} \\
\end{tabular}
\caption[]{Predicted distributions of $\delta a_\mu$ for positive (left) and negative
  (right) values of this observable.  Due to the large range of possible values for 
  $\delta a_\mu$ in these models, the positive values are shown on a log scale,
  while negative values are displayed on a linear scale.   Models surviving
  current (model set C) and future (model set D) low-energy constraints are shown
  in blue and green, respectively, assuming the future results agree with the SM, \ie, $\delta a_\mu
  = 0$. The vertical line indicates the expected bound on $\delta a_\mu$
  under this assumption, and shows that this measurement would not exclude models beyond
  the search reach of other future low-energy processes.
\label{fig:adm}}
\end{center}
\end{figure}

We also consider the pMSSM contributions to the anomalous magnetic moment 
of the muon, $\delta a_\mu$.  As is well-known, there is presently a claim \cite{Bennett:2006fi} of an observation 
of a $>3\sigma$ deviation in 
this observable from the current SM prediction.  If this deviation were to be confirmed by future experimental results, 
it would impose a particularly strong constraint on the models considered in this work.  Since there remains 
some controversy surrounding the theoretical calculation, and the measurement itself has yet to be independently 
confirmed, we have not imposed cuts on $\delta a_\mu$ in applying current low-energy constraints to our 
models (model set C in particular). However, we note that if the current experimental result for $\delta a_\mu$ 
were to be applied at the $2\sigma$ level, only 952 models in model set C would survive.  In addition, 
we have projected the impact of upcoming low energy measurements,
assuming that the SM value of $\delta a_\mu$ will be confirmed in the
future (model set D).  It is also interesting to examine the 
power of a future $\delta a_\mu$ measurement which is in agreement with the existing experimental result, 
{\it e.g.}, the same central value but with the smaller error bars that may arise from the Muon $g-2$ collaboration.  
If the current measurement of $\delta a_\mu$ persists but with the errors reduced as projected by the 
Muon $g-2$ collaboration, then none of the models in model set D would remain viable.  On the other hand, if the 
same future result is consistent with the SM, then the measurement of $\delta a_\mu$ would not provide 
any exclusionary power beyond the other low energy measurements. In other words, all models that would 
remain viable  from other future low energy flavor and CP-violation experiments are consistent 
with the SM prediction of $\delta a_\mu$ within the future anticipated sensitivity of the Muon $g-2$ experiment.  
These results are shown explicitly in Figure \ref{fig:adm}, displaying the distribution of $\delta a_\mu$ in 
the CP-violating pMSSM model sets consistent with the current and expected future low-energy constraints. Note that the 
current $2\sigma$ {\it lower} limit on $\delta a_\mu$ is $11.6 \times 10^{-10}$ since it differs from zero 
by over $3\sigma$.

In Table \ref{tab:survive} we compile
the fraction of models from set B (CP-violating) that are projected to remain viable after the most sensitive future low-energy measurements are completed, as well as the fraction of models from set A (CP-conserving) for which at least one correlated model in set B is projected to satisfy the constraints.
Clearly the neutron and electron EDMs provide the best sensitivity (strongest constraints or larger discovery reach) to the presence of phases in the CP-violating pMSSM, while other observables 
also provide good sensitivity to these new physics effects.  For the case of $(g-2)_\mu$, we assume that future experiment will uphold the current central value.  In addition, we show the fraction of models that survive after the combination of all future low-energy measurements (neglecting $(g-2)_\mu$) are performed.  This corresponds to only $\sim$0.4\% of the original $10^6$  
models generated, demonstrating the power of low-energy observables in searching for Supersymmetric CP-violating effects.

\begin{table}[!h]
\begin{center}
\begin{tabular}{c c c}
\hline
Observable & Fraction of B Models & Fraction of A Models  \\
\hline
$d_e$ & 0.00645 & 0.746 \\
$d_n$ & 0.0539 & 0.764 \\
${\rm Br}(B^+ \to X_s \gamma)$ & 0.144  & 0.877 \\
${\rm Br}(B_s \to \mu^+ \mu^-)$ & 0.148 &  0.888 \\
$(g-2)_\mu$ & 0.000366 & 0.009 \\
\hline
All future & 0.00371  &  0.587 \\
\hline
\end{tabular}
\caption[]{The fraction of models in set B (CP-violating) that are projected to
  remain viable after the most sensitive future low-energy measurements are performed, as well as the
  fraction of models in set A (CP-conserving) for which at least one correlated model in set B 
  is projected to satisfy constraints.  The $(g-2)_\mu$
  entries assume that the currently observed central value 
  continues to hold while the corresponding uncertainties shrink as projected
  in \cite{Hewett:2012ns}.  The $(g-2)_\mu$ constraint is not included in the
  ``All future'' row. 
\label{tab:survive}}
\end{center}
\end{table}

\section{Discussion and Conclusions}
\label{sec:conclusion}

This study explored the complementarity of indirect low-energy probes of the MSSM with
those arising from direct SUSY searches at the LHC. The broad suite of future low-energy
experiments examined here include flavor-violating decays of the $K$ and $B_{d,s}$ mesons - with and without CP-violating observables, meson mixing, the $g-2$ of the muon, and electric dipole moments.
To perform this investigation we employed a previously examined set 
of $10^3$ CP-conserving pMSSM models (described by specific values of the 19 pMSSM 
parameters) that satisfy existing experimental constraints, including the value of 
the observed Higgs boson mass and the direct LHC SUSY searches.  For each of these thousand models, 
we generated $10^3$ sets of CP-violating phases, corresponding to those 
associated with the parameters $M_{1,2}, \mu$ and $A_{t,b,\tau}$, thus yielding a total of $10^6$ 
models that we then employ in our study. This large set of models was then divided into various useful subsets 
depending upon how a given model is expected to respond to the anticipated 13 TeV LHC direct SUSY searches, as well as both the 
present and future set of flavor and CP-violating measurements. The sensitivity to the different types of 
experiments can then be contrasted and compared with these various subsets.        

The main result from this study is that we have shown in a quantitative manner that direct SUSY searches at the LHC and 
measurement of the low-energy observables  
considered here are almost orthogonal in their sensitivity, and probe the CP-violating pMSSM parameter space 
in completely different ways. In particular, we find that the low-energy measurements can probe pMSSM models that lie 
outside of the range of the 13 TeV LHC, even when the CP-violating phases take on modest values. However, we also find regions of 
parameter space where large phases remain allowed due to a cancellation between the contributions to CP-violating observables, 
particularly the electron and neutron EDMs. Furthermore, we have also found that the added flexibility derived 
from the inclusion of the phases can bring the pMSSM predictions closer to the observed experimental results, and 
Standard Model expectations, for rare processes such as $B_s \to \mu^+\mu^-$ which are seen to be highly restrictive in the absence of such phases. 

Of course the true impact of future low-energy observables will be highly correlated with the actual measured values;  
this is most obvious in the cases of the electron and neutron EDMs and the $g-2$ of the muon. In the case of the EDMs, 
a null measurement will severely constrain the CP violating pMSSM parameter space, especially if the anticipated 
sensitivity of future measurements is reached. On the other hand, a measurement of $g-2$ with the presently observed 
central value, but with significantly reduced experimental (and theoretical) errors, would be extremely constraining on 
this parameter space independently of what is found by the direct SUSY searches at the LHC. However, if the SM prediction for 
$g-2$ is obtained with smaller errors, then the impact on the pMSSM parameter space will turn out to be rather modest.

Although constrained by existing searches, the viable SUSY parameter space still remains quite large. Hopefully multiple 
signals for Supersymmetry will be discovered by a diverse set of experiments during the next few years.

\section*{Acknowledgements}
We are grateful for the help of Janusz Rosiek for updating the
\verb+SUSY_FLAVOR+ code.  This work was supported by the Department of
Energy, Contracts DE-AC02-06CH11357, DE-AC02-76SF00515 
and DE-FG02-12ER41811. DG has received support from the European Research Council under the European 
Union's Seventh Framework Programme (FP/2007-2013) / ERC Grant Agreement no. 279972. DG would also 
like to gratefully acknowledge the hospitality and support by the SLAC Theory Group where part of 
this work was done.  
\providecommand{\href}[2]{#2}\begingroup\raggedright\endgroup

\end{document}